**Physics-informed Machine Learning Prediction of Hubbard Interaction Parameters**

Jiyeon Kim[1*], Indukuru Ramesh Reddy[2*], Bongjae Kim[2†], Sooran Kim[1,2,3†]

[1]Department of Physics Education, Kyungpook National University, Daegu 41566, South Korea

[2]Department of Physics, Kyungpook National University, Daegu 41566, South Korea

[3]KNU G-LAMP Project Group, KNU Institute of Basic Sciences, Kyungpook National University, Daegu, 41566, Korea.

*J.K. and I.R.R. contributed equally to this work.

†Corresponding authors: bongjae@knu.ac.kr, sooran@knu.ac.kr

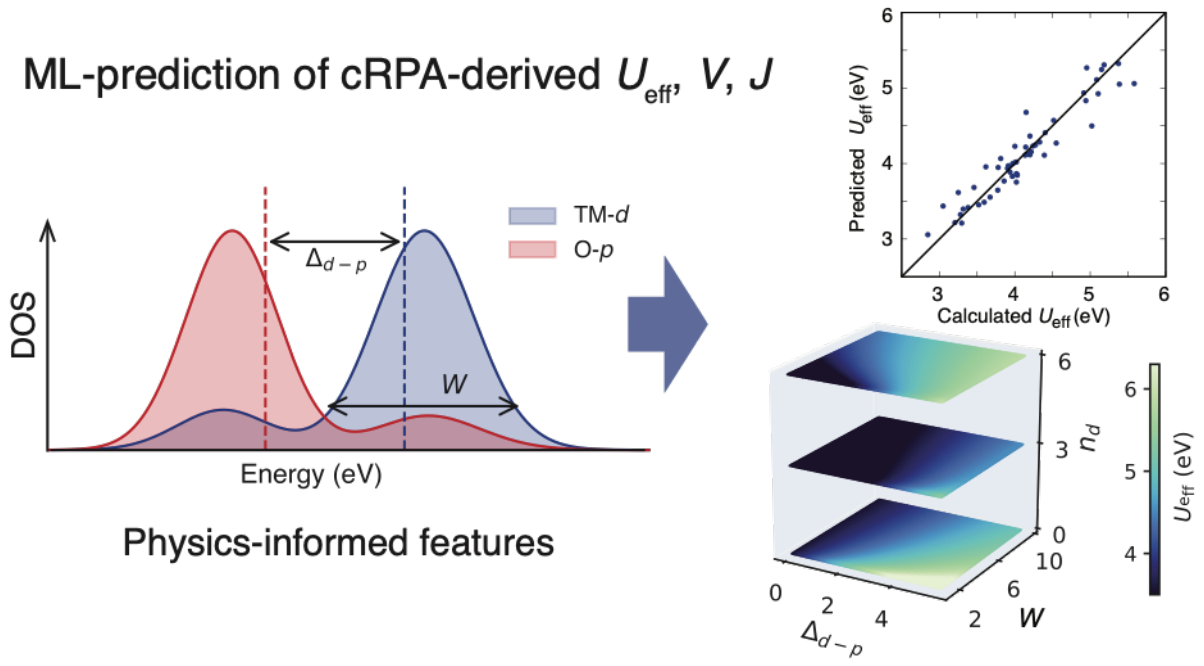

## ABSTRACT

Accurate determination of Hubbard interaction parameters is essential for beyond-DFT approaches such as DFT+$U$, DFT+DMFT, and DFT+$U$+$V$ in correlated materials. In practice, however, these parameters are often chosen empirically, limiting their transferability across materials. Advanced computational approaches such as the constrained random-phase approximation (cRPA) provide a rigorous route for evaluating Hubbard interactions, but their computational cost remains a bottleneck for large-scale materials screening. Here, we present machine-learning (ML) models for predicting cRPA-derived Hubbard interaction parameters: effective on-site $U_{\mathrm{eff}}$, inter-site $V$, and Hund's coupling $J$ for transition-metal oxides (TMOs). We combine ensemble-learning models with a regression-based brute-force search (BFS) approach to achieve both predictive accuracy and explicit analytical expressions. We construct features that capture electronic, structural, and atomic properties, including the TM-$d$ bandwidth and TM-$d$/O-$p$ band-center separation, as physically motivated descriptors of localization and screening. Our ensemble models achieve RMSEs of 0.148 eV, 0.062 eV, and 0.007 eV for $U_{\mathrm{eff}}$, $V$, and $J$, respectively. The derived analytical forms directly relate $U_{\mathrm{eff}}$ to electron localization and TM-$d$/O-$p$ hybridization, suggest the importance of hybridization and structural compactness in determining $V$, and indicate that $J$ is governed primarily by elemental descriptors of the TM ion. Together, the present study provides an efficient approach for predicting cRPA-derived $U_{\mathrm{eff}}$, $V$, and $J$, while offering physical insight into the factors underlying these Hubbard interactions.

## 1. INTRODUCTION

Strong electron correlation effects associated with transition-metal (TM) ions critically influence the electronic, structural, and magnetic properties of functional materials. To account for these correlation effects, first-principles approaches based on density functional theory (DFT) have been developed to address self-interaction errors and improve the description of localized electrons.[1–5] Among these methods, the DFT+$U$ scheme has been extensively employed to describe correlated systems, as it explicitly treats on-site Coulomb interactions ($U$) without demanding high computational cost.[6–12] However, in some correlated materials, Coulomb interactions exhibit a moderate decay with interatomic distance, highlighting the importance of long-range Coulomb interactions.[13–17] In such cases, the DFT+$U$+$V$ formalism extends DFT+$U$ by explicitly including inter-site Coulomb interactions ($V$), providing a more appropriate framework for describing these systems.[18–20]

In practice, the Hubbard parameters $U$ and/or $V$ are frequently tuned to reproduce experimental observables, such as band gaps, magnetic moments, or structural properties.[21–25] However, such empirical parameterization may limit the transferability of the DFT+$U$(+$V$) approach. To overcome this limitation, several first-principles approaches have been proposed to determine Hubbard interaction parameters.[26–30] Among them, the constrained random-phase approximation (cRPA) has been widely employed as a first-principles method[14,31–35] for determining Hubbard interaction parameters by explicitly treating electronic screening effects.[36] As a result, cRPA-derived interaction parameters are expected to show improved transferability across different materials and conditions without reliance on experimental inputs.

On top of a transparent connection to low-energy model approaches, cRPA enables analyses of individual screening channels and dynamical, frequency-dependent features in correlated-electron systems.[37–39] Furthermore, previous cRPA studies have shown that the local Hubbard interactions critically depend on a competition between orbital localization and electronic screening associated with TM-*d*/O-*p* hybridization.[14,35,40–43] Such microscopic insight into Hubbard interaction parameters is a key merit of cRPA. Nevertheless, the high computational cost of cRPA remains a major challenge for large-scale computational screening.

Recent studies have therefore explored machine-learning (ML) approaches for the efficient prediction of Hubbard interaction parameters.[44–48] Early studies employed Bayesian optimization to determine effective on-site $U$ or both on-site $U$ and inter-site $V$ parameters by fitting DFT+$U$(+$V$) band structures to hybrid band structures.[44,45] Cai *et al*. extended this approach by first determining $U$ values using Bayesian optimization and then developing random forest regression models for Mn oxides. Using structural descriptors, they identified the Mn-O bond lengths as a key factor governing the Hubbard $U$ values.[46] On the other hand, Xia *et al*. calculated $U$ values of Fe oxides using linear-response constrained DFT and developed XGBoost regression models with structural fingerprints, specifically Voronoi fingerprints, to describe the local atomic environment.[47] More recently, Uhrin *et al*. employed equivariant neural networks and predicted self-consistent $U$ and $V$ parameters obtained from linear-response density-functional perturbation theory (DFPT). They used atomic occupation matrices as descriptors that reflect geometry, local chemical environment, and electronic structure.[48]

While these studies demonstrate the effectiveness of ML in predicting Hubbard interaction parameters, ML prediction of cRPA-derived $U_{\mathrm{eff}}$, $V$, and $J$ using descriptors that capture electron localization and screening effects has remained largely unexplored. Alongside recent efforts to compare and reconcile different methodologies for calculating Hubbard parameters,[49,50] ML analysis of cRPA data may also provide insight into DFT-based approaches for estimating the Hubbard interactions, particularly in their extension to low-energy model descriptions.[27,30,51,52]

In this work, we develop ML models to predict cRPA-derived Hubbard interaction parameters: effective on-site $U_{\mathrm{eff}}$, inter-site $V$, and the Hund's coupling $J$ for a diverse set of transition-metal oxides (TMOs). Using cRPA values as reference targets, this approach enables efficient estimation of first-principles Hubbard interactions and can accelerate DFT+$U$ and DFT+$U$+$V$ workflows while retaining a physical basis. Starting from cRPA calculations of $U_{\mathrm{eff}}$, $V$, and $J$, we construct a dataset of TMOs and define descriptors associated with electronic properties, local structure, and atomic characteristics. In particular, we include electronic-structure descriptors associated with the competing effects of localization and screening. To achieve both predictive accuracy and physical interpretability, we combine two complementary approaches: ensemble learning models and a regression-based brute-force search (BFS) method. The former identifies key features governing each interaction parameter, while the latter provides explicit analytical formulas that improve model interpretability. By addressing $U_{\mathrm{eff}}$, $V$, and $J$ within a common framework, this approach enables a systematic comparison of the physical factors associated with on-site Coulomb interaction, inter-site Coulomb interaction, and Hund's exchange interaction. The overall workflow of the present study is summarized in Figure 1.

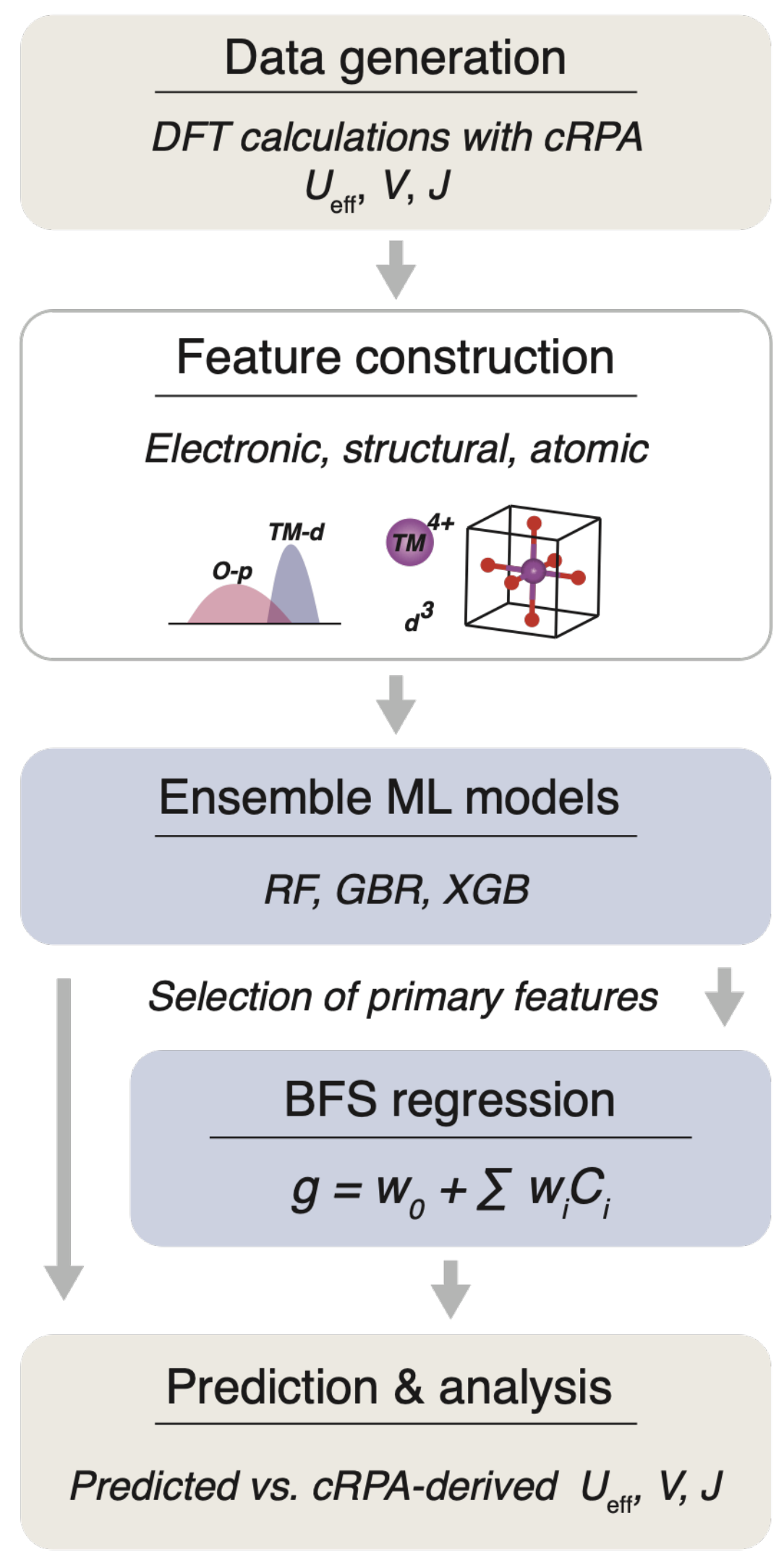

**Figure 1.** Workflow of the ML approach for predicting cRPA-derived $U_{eff}$, $V$, and $J$.

## 2. METHODS

### 2.1 cRPA Calculations

The effective on-site Coulomb interaction $U_{eff}$, inter-site Coulomb interaction $V$, and Hund's coupling $J$ for the TM-$d$ manifold were computed using cRPA. The correlated subspace, corresponding to the TM-$d$ manifold, was constructed by projecting DFT Kohn-Sham states onto Wannier bands, while the computational details of the DFT calculations and Wannierization procedure are provided in the Supporting Information. Using the resulting Wannier bands, the cRPA interaction parameters were evaluated by excluding screening processes associated with the TM-*d* bands within the correlated subspace. Specifically, the polarization contribution of the correlated subspace ($P^c$) was removed from the total polarization ($P$) to obtain the rest-space polarization, ($P^r = P - P^c$). Among the various cRPA schemes for evaluating $P^c$, we adopted the projector-based cRPA method[53] implemented within VASP.[54] Further theoretical background and technical details of the cRPA method can be found in previous studies.[34,55]

The screened Coulomb interaction can subsequently be determined by solving the following Bethe-Salpeter-type equation, $U^{-1} = [U^{bare}]^{-1} - P^r$, where $U$ ($U^{bare}$) denotes the screened (bare) Coulomb interaction. The interaction matrix elements were then evaluated in the static limit $U(\omega=0)$ among the related Wannier bands in the correlated subspace as follows:

$$\boldsymbol{U}_{ijkl}(\omega = 0) = \int\int d^3\boldsymbol{r} d^3\boldsymbol{r}' \mathcal{W}_i^*(\boldsymbol{r}) \mathcal{W}_k^*(\boldsymbol{r}') \boldsymbol{U}(\boldsymbol{r}, \boldsymbol{r}') \mathcal{W}_j(\boldsymbol{r}) \mathcal{W}_l(\boldsymbol{r}') \quad (1)$$

where $\mathcal{W}(\boldsymbol{r})$ represents the maximally localized Wannier functions spanning the correlated subspace. The on-site Coulomb interaction ($U$) and Hund's coupling ($J$) were obtained by averaging the matrix elements $\boldsymbol{U}_{iiii}$ and $\boldsymbol{U}_{ijij}$ ($i \neq j$), respectively. For the on-site $U$, all $\mathcal{W}(\boldsymbol{r})$ involved in the matrix element are centered on the same TM site. For the inter-site $V$, $\mathcal{W}_i(\boldsymbol{r})$ and $\mathcal{W}_k(\boldsymbol{r})$ are centered on one TM site, while $\mathcal{W}_j(\boldsymbol{r})$ and $\mathcal{W}_l(\boldsymbol{r})$ are centered on another TM site.

### 2.2 ML Algorithms

We employed Random Forest Regression (RF), Gradient Boosting Regression (GBR), and Extreme Gradient Boosting (XGB) models implemented using the scikit-learn Python library to predict the cRPA-derived $U_{\mathrm{eff}}$, $V$, and $J$ parameters.[56–58] Model performance was evaluated using leave-one-out cross-validation (LOOCV), and the prediction performance was quantified using the root-mean-square error (RMSE) and the coefficient of determination ($R^2$), with $R^2$ evaluated from the aggregated LOOCV predictions.

We further applied a regression-based BFS algorithm[59–62] to identify explicit relationships between the primary features and the target parameters $U_{\mathrm{eff}}$, $V$, and $J$. The primary features were selected from key descriptors through the ensemble-model analysis. Each primary feature was transformed using a predefined set of 15 nonlinear functions: 1, $x$, $1/x$, $x^2$, $1/x^2$, $x^3$, $1/x^3$, $\sqrt{x}$, $1/\sqrt{x}$, $e^x$, $e^{-x}$, $e^{1/x}$, $e^{-1/x}$, $\ln(x + 1)$, and $1/\ln(x + 1)$. These transformed variables constitute prototypical features, which were then multiplicatively combined to generate compound features. The resulting compound features were used to construct linear regression models of the form $g = w_0 + \sum_{i=1}^{m} w_i \, C_i$, where $C_i$ denotes a compound feature, $w_i$ is the corresponding regression coefficient, and $m$ is the number of compound features included in the model. Here, $g$ represents the interaction parameters: $U_{\mathrm{eff}}$, $V$, or $J$. We denote these models as $m$C$n$F, where $m$ is the number of compound features and $n$ is the number of primary features constituting each compound feature. For a given $m$C$n$F configuration, all possible combinations of compound features were first screened by fitting the full dataset, after which the top-three performing candidate models were evaluated using LOOCV.

# 3. RESULTS

## 3.1 Feature Construction

We first constructed a dataset of cRPA-derived Hubbard interaction parameters ($U_{\mathrm{eff}}$, $V$, and $J$) for TMOs with perovskite, olivine, spinel, layered, and layered perovskite structures, as listed in Table S1. To identify physics-informed features, we focus on two key factors expected to influence these interactions in TMOs: the localization of TM-$d$ orbitals and the hybridization between TM-$d$ and O-$p$ states. Based on this physical picture, we selected two electronic-structure features, the bandwidth of the TM-$d$ band ($W$) and the energy separation between the TM-$d$ and O-$p$ band centers ($\Delta_{d\text{-}p}$), as illustrated in Fig. 2(a). These two features quantify the strength of electron localization and $p$-$d$ hybridization, respectively: a larger $W$ indicates weaker localization, whereas a larger $\Delta_{d\text{-}p}$ corresponds to weaker $p$-$d$ hybridization. Details on the calculation of $W$ and $\Delta_{d\text{-}p}$ are provided in the Supporting Information.

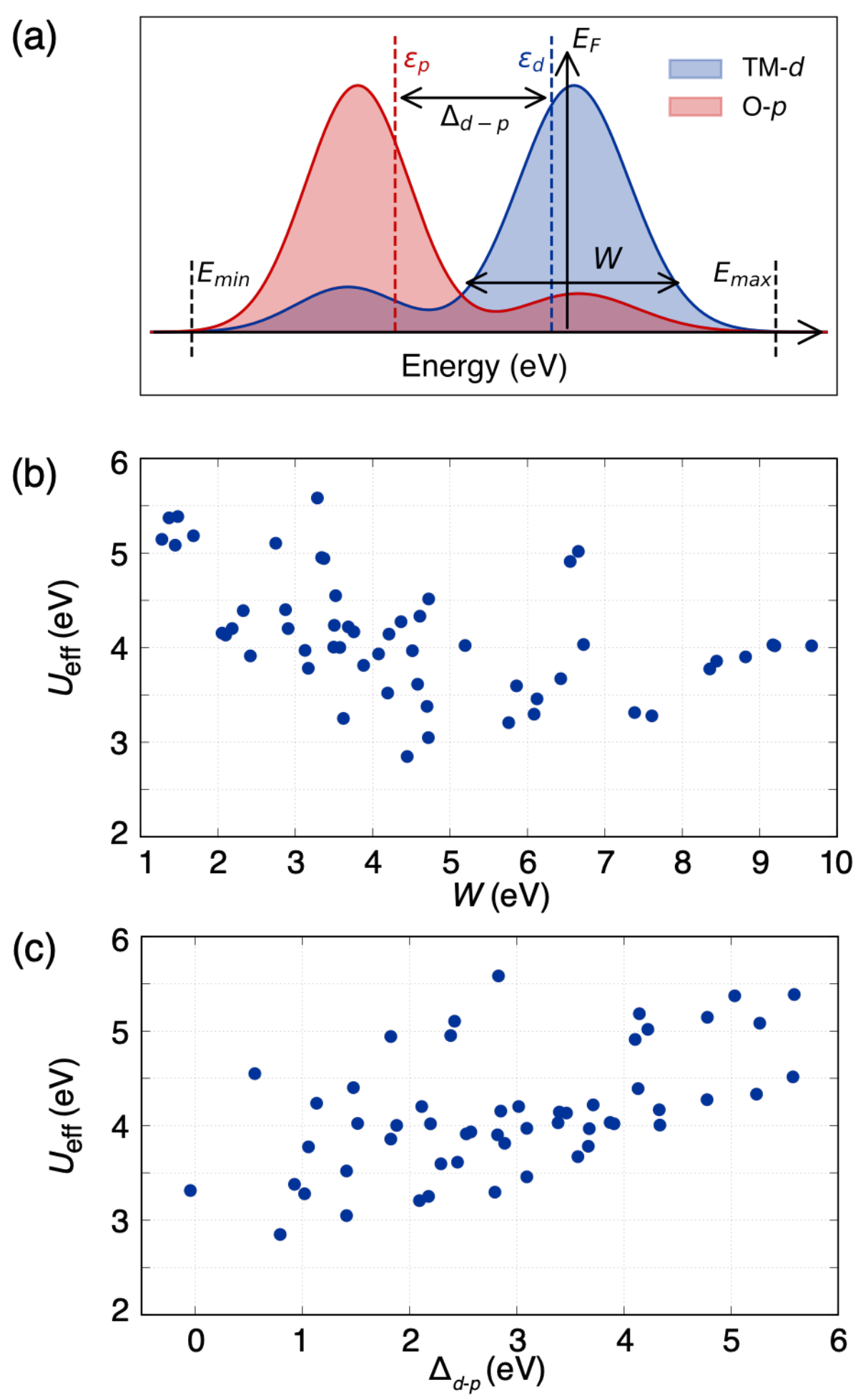

**Figure 2.** Definition of electronic descriptors and their correlations with $U_{\text{eff}}$. (a) Schematic illustration of the projected density of states (PDOS) used to define the electronic descriptors. The TM-$d$ band and O-$p$ band centers, $\boldsymbol{\varepsilon}_d$ and $\boldsymbol{\varepsilon}_p$, are computed within the energy window from $E_{\min}$ to $E_{max}$. $W$ denotes the TM-$d$ bandwidth, and $\Delta_{d\text{-}p}$ denotes the band-center separation. Correlations of $U_{eff}$ with (b) $W$ and (c) $\Delta_{d\text{-}p}$.

As shown in Fig. 2(b), the TM-$d$ bandwidth $W$ exhibits a negative correlation with $U_{eff}$. At larger bandwidths (>7 eV), a slight deviation from the overall trend is observed, which is associated with 4$d$ materials. In contrast, the band-center separation between TM-$d$ and O-$p$, $\Delta_{d\text{-}p}$, shows a positive correlation with $U_{eff}$, as in Fig. 2(c). These trends are consistent with previous cRPA studies,[14,35,41,43] in which an increase in $U$ was attributed to enhanced localization of TM-$d$ orbitals (smaller $W$) and reduced hybridization between TM-$d$ and O-$p$ orbitals (larger $\Delta_{d\text{-}p}$).

In addition to these two electronic-structure features, we also considered fundamental atomic properties (atomic number, group, and period of TM ion), compositional parameters (average $d$-electron number, average oxidation state, and the number of atoms and TM ions per unit cell), and structural characteristics (shortest TM-O bond length, volume per formula unit, and space group number), as summarized in Table 1. Among these features, the average oxidation state of the TM ion, $n_{ox}$, exhibits the strongest Pearson correlation coefficient (PCC) with $U_{eff}$ (-0.707) as shown in Table 1. This is consistent with previous observations of a correlation between Hubbard $U$ and the oxidation state of the TM ion.[63] The two electronic features discussed above also show relatively strong correlations with $U_{eff}$: $\Delta_{d\text{-}p}$ exhibits the second-highest correlation (0.539), and $W$ shows a sizable negative correlation (-0.448). The PCCs of all features with $V$ and $J$ are listed in Table S2.

**Table 1.** List of the 12 features and their Pearson correlation coefficients (PCCs) with $U_{eff}$.

| Symbol | Description | PCC |
|---|---|---|
| $W$ | Bandwidth of TM-$d$ band | -0.448 |
| $\Delta_{d\text{-}p}$ | Separation of TM-$d$ and O-$p$ band centers | 0.539 |
| $Z_{TM}$ | Atomic number of TM | -0.219 |
| $G_{TM}$ | Group number of TM | 0.028 |
| $P_{TM}$ | Period number of TM | -0.234 |
| $n_d$ | Average number of $d$ electrons of TM ion | 0.271 |
| $n_{ox}$ | Average oxidation state of TM ion | -0.707 |
| $N_{u.c}$ | Number of atoms in the unit cell | 0.459 |
| $TM_{u.c}$ | Number of TM ions in the unit cell | 0.438 |
| $d_{TM\text{-}O}$ | Shortest distance between TM and O | 0.461 |
| $V_{f.u}$ | Volume per formula unit | -0.178 |
| SG | Space group number | -0.329 |

### 3.2 Prediction of on-site $U_{eff}$

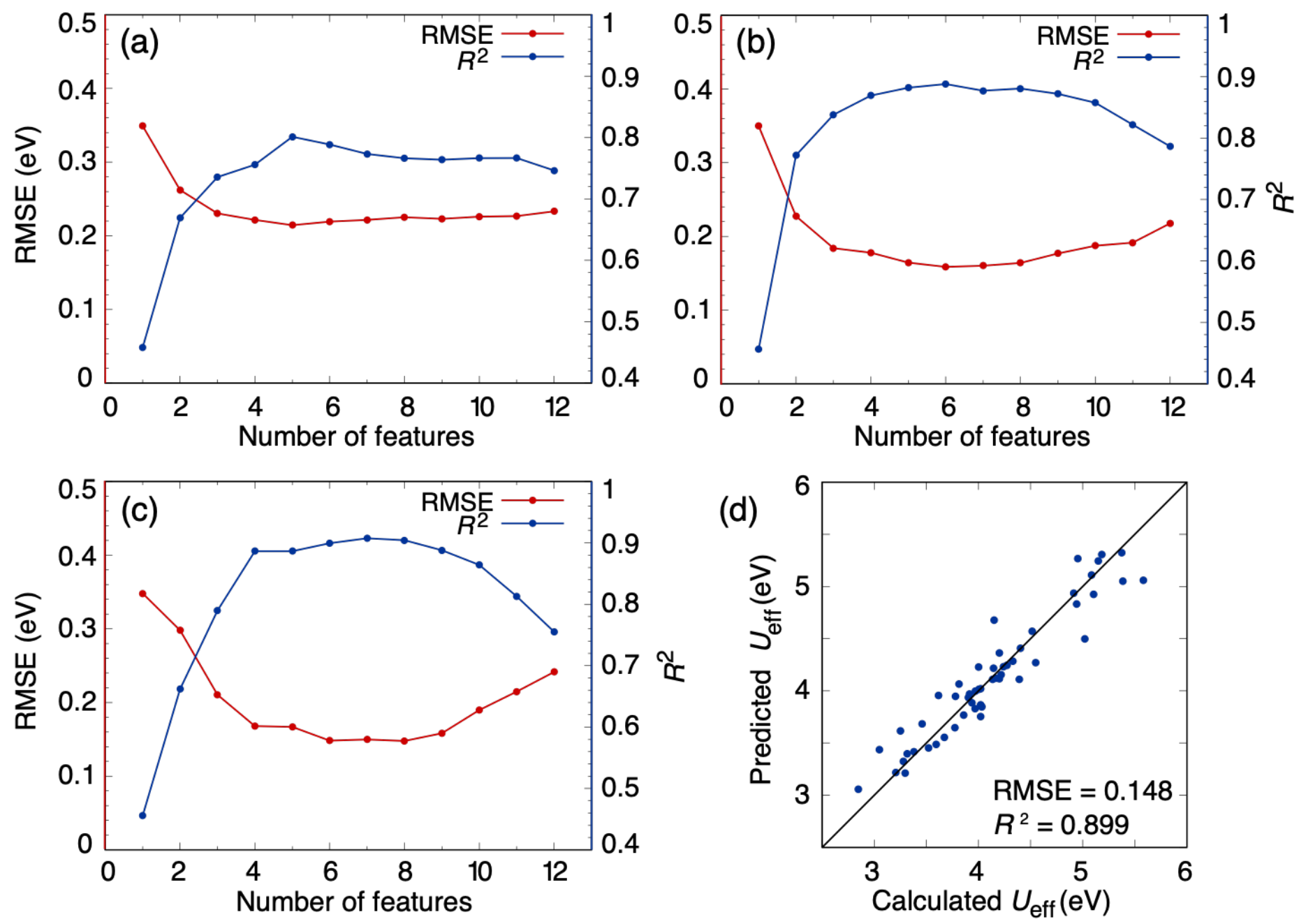

**Figure 3.** Prediction performance of ensemble-learning models for $U_{\text{eff}}$. (a) RF, (b) GBR, and (c) XGB evaluated using RMSE and $R^2$ as a function of the number of selected features. (d) Parity plot for the best-performing XGB model, comparing predicted and cRPA-derived $U_{\text{eff}}$ values.

Figure 3 compares the prediction performance of three ensemble regression models, RF, GBR, and XGB. All possible combinations of the 12 features were examined for different numbers of selected features, resulting in a total of $\Sigma\binom{12}{n} = 4{,}095$ models for each regression model. Only the best-performing combination for each number of features is shown in Fig. 3(a-c). Among the three models, the RF model achieved its best performance with an RMSE of 0.214 eV and an $R^2$ of 0.801 using five features ($n_{\text{ox}}$, $n_d$, $N_{\text{u.c}}$, $V_{\text{f.u}}$, $Z_{\text{TM}}$). The GBR model showed improved accuracy, achieving an RMSE of 0.158 eV and an $R^2$ of 0.888 with six features ($n_{\text{ox}}$, $n_d$, $N_{\text{u.c}}$, $V_{\text{f.u}}$, $TM_{\text{u.c}}$, $P_{\text{TM}}$). The highest overall performance was obtained with XGB, which achieved an RMSE of 0.148 eV and an $R^2$ of 0.899 using six features ($n_{\text{ox}}$, $n_d$, $N_{\text{u.c}}$, $V_{\text{f.u}}$, $Z_{\text{TM}}$, $G_{\text{TM}}$). Figure 3(d) presents the comparison between the predicted $U_{\text{eff}}$ and the cRPA-derived $U_{\text{eff}}$ values for the corresponding best-performing XGB model.

Across the best-performing feature combinations of all three models, four features: $n_d$, $n_{\text{ox}}$, $N_{\text{u.c}}$, and $V_{\text{f.u}}$, were consistently included. It is worth noting that these four features are readily available and do not require additional DFT calculations. This four-feature combination also corresponds to the optimal four-feature set in the XGB model, highlighting its importance for accurate predictions. For this four-feature XGB model, an RMSE of 0.168 eV and an $R^2$ of 0.886 were obtained, and the XGB performance is nearly saturated with only four features as shown in Fig. 3(c). The Gini importance (GI) from the RF algorithm further shows that $n_d$ and $n_{\text{ox}}$ make the dominant contributions to the prediction accuracy, with GI values of approximately 0.3, followed by $V_{f.u}$ (~0.2) and $N_{u.c}$ (0.15), as shown in Table S3.

We also examined the best-performing four-feature combination from the XGB model that explicitly includes both $W$ and $\Delta_{d\text{-}p}$, two physical quantities highlighted in previous cRPA studies.[14,35,41,43] For this feature set ($W$, $\Delta_{d\text{-}p}$, $n_d$, and $V_{f.u}$), the XGB model achieves a moderate prediction accuracy with an RMSE of 0.250 eV and an $R^2$ of 0.756. In this case, $\Delta_{d\text{-}p}$ and $W$ are the most important features, each with GI values of ~0.3, followed by $n_d$ (~0.25) and $V_{f.u}$ (0.16) as shown in Table S3.

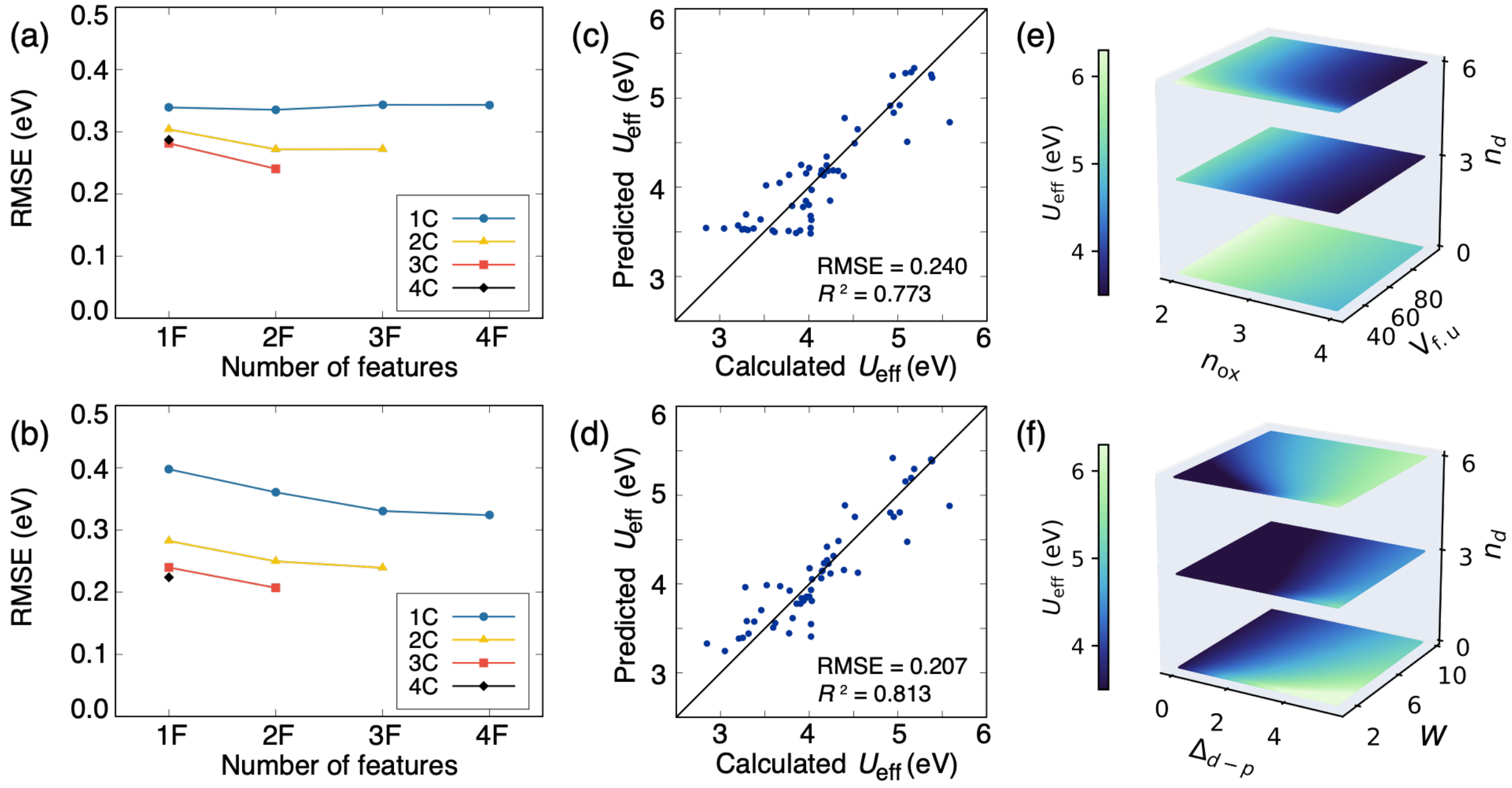

**Figure 4.** Prediction performance of the BFS models for $U_{eff}$ using two feature sets. (a,b) RMSE as a function of the number of primary and compound features. (c,d) Parity plots for the best-performing 3C2F models, comparing predicted $U_{eff}$ values with cRPA-derived values. (e,f) Three-dimensional (3D) surfaces of the corresponding 3C2F models. Panels (a), (c), and (e) correspond to feature set 1, while panels (b), (d), and (f) correspond to feature set 2.

Based on these results, we selected two four-feature sets for the BFS models. The first feature set (set 1), $n_{\mathrm{ox}}$, $n_d$, $\mathrm{N_{u.c}}$, and $\mathrm{V_{f.u}}$, was chosen as the best four-feature combination from the XGB model and includes compositional and structural descriptors. The second set (set 2), consisting of $W$, $\Delta_{d\text{-}p}$, $n_d$, and $\mathrm{V_{f.u}}$, was selected as a physically motivated set to enable a more direct analysis of the roles of electron localization and hybridization in determining the Hubbard $U$ parameter.

For each of these two feature sets, the BFS model was developed using four primary features and 15 predefined functional forms, generating 57 prototypical features and 50,625 compound features. Figures 4(a,b) show the performance of the BFS models in predicting $U_{\mathrm{eff}}$ using feature sets 1 and 2 as a function of the number of primary and compound features, denoted by the *m*C*n*F notation defined in the Methods section. For both feature sets, the prediction accuracy generally improves as the number of features increases. We consider up to 4F for 1C, 3F for 2C, and 2F for 3C, as higher-order combinations provide only marginal additional improvements. Among the evaluated configurations, the 3C2F model performed best, achieving RMSE/$R^2$ values of 0.240 eV/0.773 for set 1 and 0.207 eV/0.813 for set 2, as shown in Fig. 4(c,d). Overall, set 2 outperformed set 1, exhibiting lower RMSE values. This suggests that, within the present BFS framework, the inclusion of $W$ and $\Delta_{d\text{-}p}$ provides a more effective basis for constructing analytical expressions for $U_{\mathrm{eff}}$.

$$U_{\mathrm{eff}} = 1.898\, e^{-\left(\frac{1}{n_{\mathrm{ox}}}+n_d\right)} - 9.978\, e^{\left(\frac{1}{\mathrm{V_{f.u}}}-\frac{1}{n_{\mathrm{ox}}}\right)} + 133.485\,\frac{n_d^3}{(\mathrm{V_{f.u}})^3} + 11.337 \qquad (2)$$

Equation 2 provides the best 3C2F model for predicting the $U_{\mathrm{eff}}$ values of TMOs using feature set 1. The equations of three best-performing 3C2F models are summarized in Table S4. For set 1, the three best models show nearly identical prediction accuracies, with RMSE of 0.240-0.243 eV and $R^2$ of 0.773-0.774, and their functional forms differ only in the term involving $n_{\mathrm{ox}}$. Notably, $\mathrm{N_{u.c}}$ is absent in the equations, suggesting that its contribution is less significant than those of the other features in set 1, consistent with the GI analysis. Furthermore, these explicit analytical expressions rely only on readily available features and can therefore be applied directly without access to the original training dataset or code.

To visualize the feature dependence more clearly, three-dimensional (3D) surfaces for the best-performing 3C2F model are shown in Fig. 4(e). For set 1, $U_{\mathrm{eff}}$ decreases with increasing $n_{\mathrm{ox}}$, while larger $V_{\mathrm{f.u}}$ tends to reduce $U_{\mathrm{eff}}$ in the large-$n_d$ regime. The dependence on $n_d$ is more complex, exhibiting a nonmonotonic trend in which $U_{\mathrm{eff}}$ initially decreases and then increases with increasing $n_d$. (See also Figure S1.)

$$U_{\mathrm{eff}} = 0.626\,\frac{\Delta_{d-p}}{\ln(1+W)} - 1.963\sqrt{\frac{n_d}{W}} + 0.275\,\frac{{n_d}^2}{\ln(1+\mathrm{V}_{\mathrm{f.u}})} + 3.559 \quad (3)$$

For feature set 2, eq 3 gives the best-performing 3C2F model. The three best-performing models exhibit RMSE values of 0.207-0.225 eV and $R^2$ values of 0.812-0.813 (Table S4). Importantly, these equations provide a direct and quantitative relation between $U_{\mathrm{eff}}$ and the electronic descriptors associated with localization and screening, represented by $W$ and $\Delta_{d\text{-}p}$, respectively.

Figure 4(f) shows the 3D plot of eq 3, illustrating how $\Delta_{d\text{-}p}$, $W$, and $n_d$ collectively determine $U_{\mathrm{eff}}$. For visualization, $V_{\mathrm{f.u}}$ was fixed at the dataset average (59.753 Å$^3$), as it has the smallest GI among set 2. As indicated by the first term in eq 3 and shown in Fig. 4(f), an increase in $\Delta_{d\text{-}p}$, which reflects reduced *d-p* hybridization, leads to a larger $U_{\mathrm{eff}}$, in agreement with the trend in Fig. 2(c). In contrast, because $W$ appears in competing terms in eq 3, its effect on $U_{\mathrm{eff}}$ is $n_d$-dependent. As a result, increasing $W$ lowers $U_{\mathrm{eff}}$ at small $n_d$, but this tendency can be weakened or reversed for larger $n_d$. The dependence on $n_d$ is also intricate, as observed for set 1, where $U_{\mathrm{eff}}$ first decreases and then increases with increasing $n_d$, as shown in Fig. S1(b). This nonmonotonic behavior of $U_{\mathrm{eff}}$ is consistent with trends reported in previous studies.[14,43,64,65] In addition, according to the third term in eq 3, increasing $V_{f.u}$ reduces $U_{eff}$. Overall, the analytical expressions for $U_{eff}$ directly connect its variation to localization and TM-*d*/O-*p* hybridization, while also capturing additional contributions from composition- and structure-related descriptors.

### 3.3 Prediction of inter-site *V*

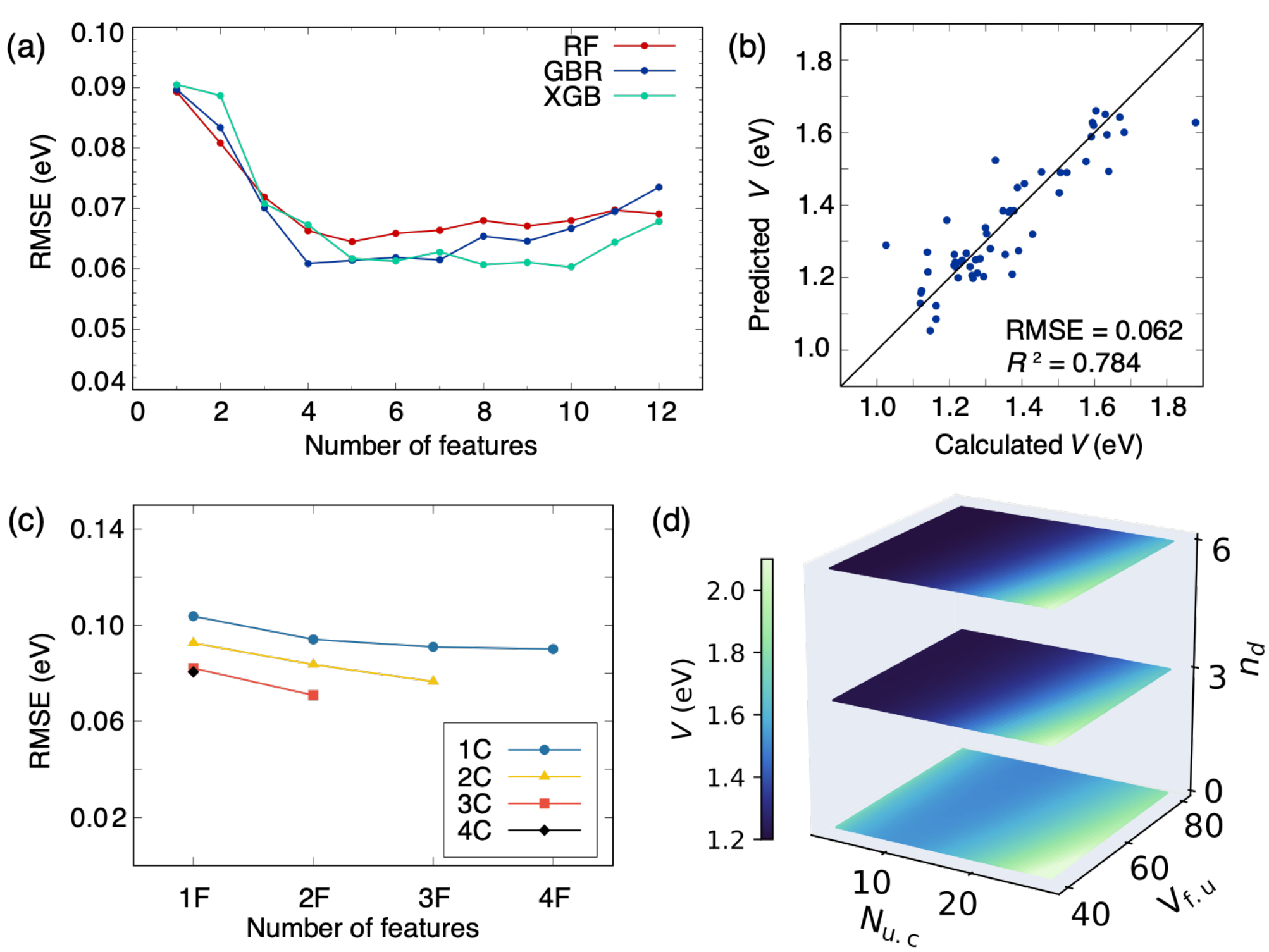

**Figure 5.** Prediction performance of ensemble-learning and BFS models for *V*. (a) Model performance of RF, GBR, and XGB models as a function of the number of selected features. (b) Parity plot for the XGB model with five selected features, comparing predicted and cRPA-derived *V* values. (c) BFS model performance using feature set 1 ($n_d$, $N_{u.c}$, $V_{f.u}$, and $P_{TM}$). (d) 3D surface of the best-performing BFS model for feature set 1.

We next extend our analysis to the prediction of *V* using the RF, GBR, and XGB models, and the corresponding performance is shown in Fig. 5(a). As the number of features increases, all three models exhibit a rapid decrease in RMSE, which saturates with four to six features. The RF model achieves its best performance with five features ($n_d$, $N_{u.c}$, $V_{f.u}$, SG, and $Z_{TM}$), giving an RMSE of 0.065 eV and an $R^2$ of 0.686. The GBR model shows improved performance, reaching an RMSE of 0.061 eV and an $R^2$ of 0.764 with four features ($n_{ox}$, $n_d$, $V_{f.u}$, and SG). For XGB, the minimum RMSE of 0.060 eV was obtained using ten features, while its performance is already close to saturation with five features ($n_d$, $N_{u.c}$, $V_{f.u}$, SG, and $Z_{TM}$), yielding an RMSE of 0.062 eV and an $R^2$ of 0.784. Figure 5(b) shows the corresponding parity plot for this five-feature XGB model.

For the BFS analysis, we focused on compact four-feature sets, since the ensemble-model performance is already nearly saturated in this regime, while fewer variables make the resulting analytical expressions more tractable and interpretable. In selecting the primary features, we considered not only predictive accuracy but also the physical transparency of the resulting formulas. Although the best four-feature GBR model showed slightly better performance, it included the space-group number (SG), whose role is less straightforward to interpret in an explicit analytical expression. We therefore adopted the best four-feature XGB combination as the first primary feature set (set 1), namely $n_d$, $N_{u.c}$, $V_{f.u}$, and $P_{TM}$. As a second primary feature set (set 2), we selected the best-performing four-feature combination that explicitly includes $\Delta_{d\text{-}p}$ and $W$, namely $\Delta_{d\text{-}p}$, $W$, $n_d$, and $TM_{u.c}$.

The prediction performance of the BFS models for set 1 and set 2 is shown in Fig. 5(c) and Fig. S2(a), respectively. The best performance of the BFS model for set 1 is obtained with the 3C2F configuration, yielding an RMSE of 0.071 eV and an $R^2$ of 0.763, while the best performance for set 2 is achieved with the 2C3F configuration, yielding an RMSE of 0.074 eV and $R^2$ of 0.697. Overall, set 1 provides better predictive accuracy than set 2.

$$V = 0.919\,\frac{e^{-n_d}}{\sqrt{\mathrm{N_{u.c}}}} + 1.673\times10^{-3}\,\frac{(\mathrm{N_{u.c}})^3}{\mathrm{V_{f.u}}} + 8.342\times10^{9}\,e^{-\mathrm{V_{f.u}}}(\mathrm{N_{u.c}})^3 + 1.193 \qquad (4)$$

Equation 4 gives the best-performing BFS model for set 1. The three best-performing models for set 1 are summarized in Table S5, and their functional forms are highly similar. In particular, only the first term associated with $\mathrm{N_{u.c}}$ differs slightly among the three models, indicating the robustness of the resulting expressions. Notably, $\mathrm{P_{TM}}$ does not appear in the final equations, suggesting that it plays a smaller role than the other descriptors in set 1. As in the $U_{\mathrm{eff}}$ case, these analytical expressions depend on features that do not require any prior calculations, which enhances their practical applicability to other TMO materials.

Figure 5(d) shows the 3D plot of the best-performing 3C2F model equation for set 1. Smaller volumes lead to larger $V$, and larger $\mathrm{N_{u.c}}$ tends to be associated with larger $V$. These results suggest that structurally compact or dense systems exhibit larger inter-site Coulomb interactions. This can be understood from the general distance dependence of inter-site Coulomb interactions, which decrease with increasing interatomic distances.[13,14,66] In addition, increasing $n_d$ leads to smaller $V$.

$$V = 2.363\times10^{-3}\left(\Delta_{d-p}\right)^3 e^{\left(\frac{1}{\mathrm{TM_{u.c}}}-n_d\right)} + 8.392\times10^{-4}\,n_d \ln\left(1+\Delta_{d-p}\right) e^{\mathrm{TM_{u.c}}} + 1.197 \qquad (5)$$

For feature set 2, eq 5 gives the best-performing model. The three best-performing models for set 2 are also summarized in Table S5, and their functional forms are again similar, with only slight variations in the first term. The dependence of $V$ on $\Delta_{d\text{-}p}$ is particularly pronounced, as evident from both eq 5 and Fig. S2(b): increasing $\Delta_{d\text{-}p}$ leads to a larger $V$. Interestingly, $W$ does not appear in the final equations, indicating that it makes the smallest contribution among the set 2 descriptors. This suggests that, in contrast to the local interaction $U_{\mathrm{eff}}$, which reflects both $d$-orbital localization and TM-$d$/O-$p$ hybridization, the nonlocal interaction $V$ is less directly associated with $d$-orbital localization and is more strongly influenced by hybridization-related screening. As shown in Fig. S2(b), a larger $\mathrm{TM_{u.c.}}$ generally increases $V$ at moderate and large $n_d$. This behavior suggests that a higher density of interacting TM ions can enhance the inter-site Coulomb interaction. An increase in $n_d$ generally decreases $V$. Taken together, the analytical expressions for $V$ suggest that both hybridization-related screening and structural compactness contribute to the inter-site interaction. This structural trend is qualitatively consistent with the Ohno-type decay with Resta-screened distribution of the Coulomb interactions.[13,14,66]

### 3.4 Prediction of Hund's coupling *J*

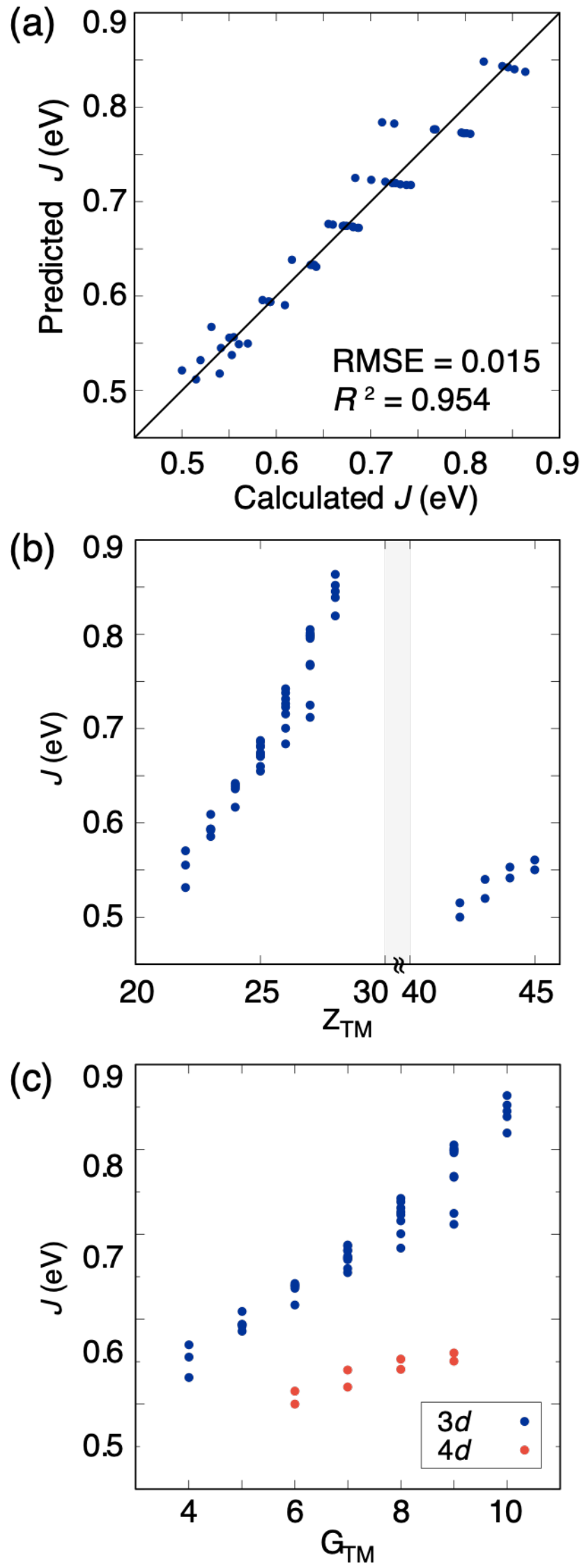

**Figure 6.** Prediction performance of the ML model for $J$ and key descriptors. (a) Parity plot for the RF model with a single feature, $Z_{TM}$, comparing predicted and cRPA-derived $J$ values. Correlations of $J$ with (b) $Z_{TM}$ and (c) $G_{TM}$.

We next turn to the prediction of the Hund's coupling parameter $J$ using the same ML framework. In contrast to $U_{eff}$ and $V$, $J$ is predicted remarkably well even with a single descriptor. As shown in Figure S3, using only $Z_{TM}$, the RF, GBR, and XGB models already achieve RMSE values of 0.015-0.016 eV and $R^2$ of 0.954. Figure 6(a) shows the corresponding parity plot from the RF model, while Fig. 6(b) shows the dependence of $J$ on $Z_{TM}$. Two distinct branches are observed, corresponding to 3*d* and 4*d* TM compounds, and within each branch $J$ increases with increasing $Z_{TM}$. This trend is consistent with the expectation that Hund's coupling is primarily determined by the local character of TM *d*-orbitals and tends to increase with *d*-electron filling.[14,34,35,42,43]

Among the three ensemble models, XGB gives the best performance, achieving an RMSE of 0.007 eV and an $R^2$ of 0.989 using three features: $G_{TM}$, $P_{TM}$, and SG. Here, $G_{TM}$ describes the sequence of TM elements within a given period, and $P_{TM}$ distinguishes the 3*d* and 4*d* series. As shown in Fig. 6(c), $J$ increases systematically with $G_{TM}$, while the 3*d* and 4*d* materials form two clearly separated series distinguished by $P_{TM}$.

We further analyzed $J$ using BFS models based on two three-feature sets selected from the XGB results: set 1 ($G_{TM}$, $P_{TM}$, and SG), and set 2 ($\Delta_{d-p}$, $W$, and $Z_{TM}$). The corresponding BFS prediction performance is summarized in Fig. S4(a,b). Since both sets already achieve $R^2 > 0.9$ with simple expressions, we selected 1C2F and 2C2F as representative forms for set 1 and set 2, respectively. The 1C2F model for set 1 gives an RMSE of 0.015 eV and an $R^2$ of 0.954, while the 2C2F model for set 2 yields an RMSE of 0.022 eV and an $R^2$ of 0.929.

Table S6 summarizes the representative BFS expressions for predicting $J$. For set 1, the best-performing BFS model is

$$J = 3.001 \frac{\mathrm{G_{TM}}}{\mathrm{P_{TM}^3}} + 0.354 \quad (6)$$

which depends only on elemental descriptors. The inverse dependence on $P_{TM}$ indicates reduced exchange interaction with increasing period of the TM series, while the increase with $G_{TM}$ is consistent with the trend in Fig. 6(c). A two-dimensional (2D) projection of eq 6 is shown in Fig. S4(c).

For set 2, the resulting BFS expressions in Table S6 retain only $W$ and $Z_{TM}$, indicating that $\Delta_{d\text{-}p}$ does not appear in the final model. As shown in Fig. S4(d,e), $J$ tends to increase with $Z_{TM}$ and generally decreases with increasing $W$. Since 4*d* compounds typically have broader *d*-bands than 3*d* compounds, $W$ effectively distinguishes the two TM series. Thus, while $J$ is governed primarily by intrinsic atomic characteristics of the TM ion, additional descriptors such as $P_{TM}$ and $W$ help account for systematic differences between TM series.

In contrast to Hubbard $U$ and inter-site $V$, Hund's $J$ is weakly screened within cRPA, so that the bare atomic trend is largely preserved, and atomic-like localization effects remain prominent. Consequently, the contraction of the *d*-orbital with increasing nuclear charge, i.e. $Z_{TM}$ (and $G_{TM}$), plays an important role in determining $J$, while screening-related parameters such as $\Delta_{d\text{-p}}$ have only a limited influence.

To further reduce the influence of the 3*d*/4*d* distinction, we additionally performed 1C1F-BFS calculations using only 3*d* compounds. Within this restricted dataset, the optimal set 1 model is $J = 0.003\ G_{TM}^2 + 0.510$, while the best set 2 model is $J = 2.5 \times 10^{-5}\ Z_{TM}^3 + 0.286$. Both models give the same prediction accuracy, with an RMSE of 0.015 eV and an $R^2$ of 0.943. These results further support that $J$ can be described accurately using a minimal set of elemental descriptors within the same TM series.

## 4. CONCLUSIONS

In this work, we developed ML models for predicting cRPA-derived Hubbard interaction parameters, $U_{\mathrm{eff}}$, $V$, and Hund's coupling $J$ in TMOs by combining ensemble-learning models with a regression-based BFS analysis. This approach achieves high predictive accuracy while also providing analytical expressions that offer insight into the underlying physics of the interaction parameters. The selected descriptors were designed to capture electronic, structural, and atomic characteristics, including physically motivated quantities associated with localization and TM-$d$/O-$p$ hybridization.

This study shows that the three interaction parameters can be interpreted within a common approach while still exhibiting distinct physical tendencies. For $U_{\mathrm{eff}}$, the ensemble models showed strong predictive performance, and the BFS analysis yielded analytical formulas that directly connect $U_{\mathrm{eff}}$ to localization and hybridization effects through descriptors such as $W$ and $\Delta_{d\text{-}p}$. For $V$, the ensemble models also achieved good predictive performance, and the derived analytical forms suggested a more prominent role of $\Delta_{d\text{-}p}$ than $W$, together with structural compactness. For $J$, accurate prediction was achieved using only a very small number of elemental descriptors, indicating that Hund's coupling is governed primarily by intrinsic atomic characteristics of the TM ion.

Overall, the present framework demonstrates that cRPA-derived Hubbard parameters can be predicted not only efficiently but also in a physically interpretable manner. More broadly, our study provides a practical route for identifying descriptor-property relationships in correlated oxides and accelerating beyond-DFT workflows through physics-informed machine learning.

## ASSOCIATED CONTENT

### Data Availability Statement

The code used in this study is available at https://github.com/CMPL-KNU/cRPA-parameter-prediction. The data supporting the findings of this study are available from the corresponding author upon reasonable request.

### Supporting Information

Additional details on the calculation of electronic descriptors and computational methods; cRPA-derived $U_{\mathrm{eff}}$, $V$, and $J$ values for transition-metal oxides; Pearson correlation coefficients and Gini importance values; brute-force search model equations; additional figures showing descriptor dependence and model performance for $U_{\mathrm{eff}}$, $V$, and $J$; energy windows for band-center calculations and Wannier-function construction; and Wannier-interpolated band structures.

## AUTHOR INFORMATION

### Corresponding Authors

**Sooran Kim** - Department of Physics Education, Kyungpook National University, Daegu 41566, South Korea; Department of Physics, Kyungpook National University, Daegu 41566, South Korea; KNU G-LAMP Project Group, KNU Institute of Basic Sciences, Kyungpook National University, Daegu, 41566, Korea.

**Bongjae Kim** - Department of Physics, Kyungpook National University, Daegu 41566, South Korea

**Authors**

**Jiyeon Kim** - Department of Physics Education, Kyungpook National University, Daegu 41566, South Korea

**Indukuru Ramesh Reddy** - Department of Physics, Kyungpook National University, Daegu 41566, South Korea

**Author contributions**

J.K. developed the ML models, and I.R.R. performed the cRPA calculations. S.K. and B.K. conceived and supervised the study. All authors contributed to data analysis and participated in writing and revising the manuscript.

**Notes**

The authors declare no competing interests.

**ACKNOWLEDGMENTS**

This work was supported by the National Research Foundation of Korea (NRF) (Grant No. 2022R1F1A1063011, RS-2025-00557388). This work was also supported by Learning & Academic research institution for Master's·PhD students, and Postdocs (LAMP) Program of the NRF grant funded by the Ministry of Education (No. RS-2023-00301914). BK was also supported by the National Research Foundation of Korea (NRF) grant funded by the Korea government (MSIT) (Grants No. RS-2023-00256050, No. RS-2026-25472078, and No. RS-2022-NR068223) and KISTI Supercomputing Center (Project No. KSC-2023-CRE-0413).

**Physics-informed Machine Learning Prediction of Hubbard Interaction Parameters**

Jiyeon Kim[1*], Indukuru Ramesh Reddy[2*], Bongjae Kim[2†], Sooran Kim[1,2,3†]

[1]Department of Physics Education, Kyungpook National University, Daegu 41566, South Korea

[2]Department of Physics, Kyungpook National University, Daegu 41566, South Korea

[3]KNU G-LAMP Project Group, KNU Institute of Basic Sciences, Kyungpook National University, Daegu, 41566, Korea.

*J.K. and I.R.R. contributed equally to this work.

†Corresponding authors: bongjae@knu.ac.kr, sooran@knu.ac.kr

## Note S1. Computational details

The DFT calculations were performed using VASP.[1] The generalized gradient approximation (GGA) in the Perdew-Burke-Ernzerhof (PBE) parametrization was used for the exchange-correlation functional.[2] The electron-ionic core interaction was represented by projector augmented wave potentials (PAWs).[3] A large plane-wave cutoff energy 500 eV was employed for all the perovskite and layered perovskite structures, while 650 eV was used for other structures. For the response function calculations, the cutoff energy was set to the default value, corresponding to two-thirds of the plane-wave cutoff energy. The Brillouin zone was sampled using Γ-centered *k*-point meshes with a reciprocal space resolution of 0.03-0.04 $\times$ 2π $Å^{-1}$. The convergence criterion for the self-consistent electronic energy minimization was set to $10^{-8}$ eV.

The interaction parameters of the TM-*d* manifold, including the effective on-site Coulomb interaction ($U_{eff}$), inter-site Coulomb interaction ($V$), and Hund's coupling ($J$), were computed using the constrained random-phase approximation (cRPA) within the projector-based implementation in VASP.[1,4] The central part of cRPA is in defining the correlated subspace, typically corresponding to the TM-*d* manifold, and obtaining the associated polarizability ($P^c$). Here, the correlated subspace was defined using maximally localized Wannier functions (MLWFs) constructed from DFT Kohn–Sham states using WANNIER90[5] interfaced with VASP via VASP2WANNIER90.[6] The MLWFs were constructed within a *d–dp* model including both TM-*d* and O-*p* states, while the interaction parameters were evaluated for the TM-*d* manifold only. The disentanglement energy windows ($\mathbb{W}$) used for constructing the Wannier functions are listed in Table S7. Default parameters were used for both the disentanglement convergence criterion and Wannierization.[7,8]

To account for $V$ along both the in-plane and out-of-plane directions, supercells were constructed according to the crystal geometry. For the perovskite and layered perovskite structures, $SrMO_3$ (M = 3*d* and 4*d*) and $Sr_2MO_4$ (M = 3*d* and 4*d*), $\sqrt{2}\times\sqrt{2}\times2$ and $\sqrt{2}\times\sqrt{2}\times1$ supercells were employed, respectively. For layered (Li/Na)$MO_2$ materials, a 1×1×2 supercell was used for the monoclinic structures (M = Mn and Ni), while a $1\times\sqrt{3}\times1$ supercell was adopted for the remaining hexagonal structures. For the olivine and spinel structures, both in-plane and out-of-plane $V$ are already contained within the 1×1×1 olivine cell and the primitive spinel cell. Thus, cRPA calculations were performed without constructing supercells for these structures to reduce computational cost.

## Note S2. Calculation of $\Delta_{d\text{-}p}$ and $W$

The TM-*d* and O-*p* band centers, $\boldsymbol{\varepsilon}_d$ and $\boldsymbol{\varepsilon}_p$, were calculated from projected density of states (PDOS)[9] as

$$\boldsymbol{\varepsilon}_{d(p)} = \frac{\int_{E\ min}^{E\ max} \boldsymbol{n}_{\boldsymbol{d}(\boldsymbol{p})}(\boldsymbol{\epsilon}_{\boldsymbol{d}(\boldsymbol{p})})\,\boldsymbol{\epsilon}_{\boldsymbol{d}(\boldsymbol{p})}\,\boldsymbol{d}\boldsymbol{\epsilon}_{\boldsymbol{d}(\boldsymbol{p})}}{\int_{E\ min}^{E\ max} \boldsymbol{n}_{\boldsymbol{d}(\boldsymbol{p})}(\boldsymbol{\epsilon}_{\boldsymbol{d}(\boldsymbol{p})})\,\boldsymbol{d}\boldsymbol{\epsilon}_{\boldsymbol{d}(\boldsymbol{p})}}$$

where $\boldsymbol{n}_{\boldsymbol{d}(\boldsymbol{p})}$ is the PDOS and $\boldsymbol{\epsilon}_{\boldsymbol{d}(\boldsymbol{p})}$ denotes the energy of the TM-*d* (O-*p*) states. The integration limits $E_{\min}$ and $E_{\max}$ define the energy window used to calculate the band centers, as illustrated schematically in Fig. 2(a) of the main text. Accordingly, $\Delta_{d\text{-}p}$ is obtained from the energy separation between $\boldsymbol{\varepsilon}_d$ and $\boldsymbol{\varepsilon}_p$. The corresponding $E_{\min}$ and $E_{\max}$ values are given in Table S7.

The TM-*d* bandwidth ($W$) was calculated from the energy width of TM-*d* Wannier bands (green dashed lines) shown in band structures presented in Figures S5-S10. The Wannier-interpolated bands and corresponding density functional theory (DFT) band structures for $SrMO_3$ (M = 3*d*) and $LiMO_2$ (M = 3*d*) have been reported previously.[7,8]

**Table S1.** List of transition-metal oxide compounds and cRPA-derived $U_{eff}$, $V$, and $J$ values.

| Structure type | List | $U_{eff}$ | $V$ | $J$ |
|---|---|---|---|---|
| Perovskite | $SrTiO_3$ | 4.912 | 1.640 | 0.570 |
| | $SrVO_3$ | 4.033 | 1.354 | 0.609 |
| | $SrCrO_3$ | 3.459 | 1.300 | 0.640 |
| | $SrMnO_3$ | 3.596 | 1.313 | 0.672 |
| | $SrFeO_3$ | 4.021 | 1.192 | 0.700 |
| | $SrCoO_3$ | 3.378 | 1.391 | 0.725 |
| | $SrMoO_3$ | 4.029 | 1.231 | 0.515 |
| | $SrTcO_3$ | 3.903 | 1.213 | 0.540 |
| | $SrRuO_3$ | 3.857 | 1.262 | 0.553 |
| | $SrRhO_3$ | 3.278 | 1.302 | 0.561 |
| Olivine | $LiMnPO_4$ | 5.388 | 1.682 | 0.670 |
| | $LiFePO_4$ | 5.374 | 1.670 | 0.723 |
| | $NaMnPO_4$ | 5.085 | 1.591 | 0.660 |
| | $NaFePO_4$ | 5.148 | 1.598 | 0.716 |
| | $NaCoPO_4$ | 5.184 | 1.604 | 0.767 |
| | $MnPO_4$ | 4.391 | 1.576 | 0.682 |
| | $FePO_4$ | 4.135 | 1.595 | 0.726 |
| | $CoPO_4$ | 4.152 | 1.630 | 0.769 |
| | $NiPO_4$ | 4.202 | 1.635 | 0.819 |
| Spinel | $LiCr_2O_4$ | 3.968 | 1.429 | 0.642 |
| | $LiMn_2O_4$ | 3.814 | 1.367 | 0.687 |
| | $LiFe_2O_4$ | 3.936 | 1.363 | 0.742 |
| | $LiCo_2O_4$ | 4.002 | 1.406 | 0.801 |
| | $LiNi_2O_4$ | 4.239 | 1.387 | 0.864 |
| | $Li_2Ti_2O_4$ | 4.516 | 1.506 | 0.531 |
| | $Li_2V_2O_4$ | 4.274 | 1.454 | 0.586 |
| | $Li_2Mn_2O_4$ | 4.143 | 1.327 | 0.672 |
| | $Li_2Co_2O_4$ | 4.956 | 1.879 | 0.798 |
| | $Mn_2O_4$ | 3.615 | 1.378 | 0.686 |
| Layered | $LiVO_2$ | 4.332 | 1.285 | 0.592 |
| | $LiCrO_2$ | 4.168 | 1.272 | 0.639 |
| | $LiMnO_2$ | 4.219 | 1.295 | 0.674 |
| | $LiFeO_2$ | 4.202 | 1.277 | 0.738 |
| | $LiCoO_2$ | 5.584 | 1.524 | 0.805 |
| | $LiNiO_2$ | 4.942 | 1.139 | 0.852 |
| | $NaVO_2$ | 4.005 | 1.121 | 0.594 |
| | $NaCrO_2$ | 3.781 | 1.120 | 0.636 |
| | $NaMnO_2$ | 3.972 | 1.265 | 0.673 |
| | $NaFeO_2$ | 3.915 | 1.140 | 0.731 |
| | $NaCoO_2$ | 5.104 | 1.347 | 0.799 |
| | $NaNiO_2$ | 4.404 | 1.163 | 0.845 |
| | $MnO_2$ | 3.253 | 1.146 | 0.681 |
| | $CoO_2$ | 3.521 | 1.025 | 0.796 |
| | $NiO_2$ | 4.550 | 1.373 | 0.839 |
| Layered perovskite | $Sr_2TiO_4$ | 5.019 | 1.503 | 0.555 |
| | $Sr_2VO_4$ | 3.673 | 1.235 | 0.594 |
| | $Sr_2CrO_4$ | 3.297 | 1.216 | 0.617 |
| | $Sr_2MnO_4$ | 3.207 | 1.217 | 0.655 |
| | $Sr_2FeO_4$ | 3.049 | 1.223 | 0.684 |
| | $Sr_2CoO_4$ | 2.849 | 1.246 | 0.712 |
| | $Sr_2MoO_4$ | 4.020 | 1.162 | 0.500 |
| | $Sr_2TcO_4$ | 4.020 | 1.123 | 0.520 |
| | $Sr_2RhO_4$ | 3.316 | 1.257 | 0.550 |
| | $Sr_2RuO_4$ | 3.773 | 1.213 | 0.541 |

**Table S2.** Pearson correlation coefficients between each feature and the interaction parameters $U_{\mathrm{eff}}$, $V$, and $J$.

| Feature | $U_{\mathrm{eff}}$ | $V$ | $J$ |
|---|---|---|---|
| $W$ | -0.448 | -0.458 | -0.731 |
| $\Delta_{d\text{-}p}$ | 0.539 | 0.405 | -0.312 |
| $Z_{\mathrm{TM}}$ | -0.219 | -0.283 | -0.394 |
| $G_{\mathrm{TM}}$ | 0.028 | 0.051 | 0.745 |
| $P_{\mathrm{TM}}$ | -0.234 | -0.307 | -0.601 |
| $n_d$ | 0.271 | 0.228 | 0.789 |
| $n_{\mathrm{ox}}$ | -0.707 | -0.526 | -0.350 |
| $N_{\mathrm{u.c}}$ | 0.459 | 0.638 | 0.145 |
| $TM_{\mathrm{u.c}}$ | 0.438 | 0.578 | 0.310 |
| $d_{\mathrm{TM\text{-}O}}$ | 0.461 | 0.184 | -0.114 |
| $V_{\mathrm{f.u}}$ | -0.178 | 0.066 | -0.374 |
| SG | -0.329 | -0.132 | -0.277 |

**Table S3.** Gini importance of features in sets 1 and 2 for predicting $U_{\mathrm{eff}}$.

| | Set 1 | GI |
|---|---|---|
| 1 | $n_d$ | 0.321 |
| 2 | $n_{\mathrm{ox}}$ | 0.290 |
| 3 | $V_{\mathrm{f.u}}$ | 0.235 |
| 4 | $N_{\mathrm{u.c}}$ | 0.154 |
| | Set 2 | GI |
| 1 | $\Delta_{d\text{-}p}$ | 0.308 |
| 2 | $W$ | 0.284 |
| 3 | $n_d$ | 0.247 |
| 4 | $V_{\mathrm{f.u}}$ | 0.161 |

**Table S4.** Equations of the best-performing 3C2F models for predicting $U_{\mathrm{eff}}$

| | Model equations (set 1) | RMSE (eV) | $R^2$ |
|---|---|---|---|
| 1 | $1.898\, e^{-(\frac{1}{n_{\mathrm{ox}}}+n_d)} - 9.978\, e^{(\frac{1}{\mathrm{V_{f.u}}}-\frac{1}{n_{\mathrm{ox}}})} + 133.485 \frac{n_d^3}{(\mathrm{V_{f.u}})^3} + 11.337$ | 0.240 | 0.773 |
| 2 | $0.741\sqrt{n_{\mathrm{ox}}}\, e^{-n_d} - 9.991\, e^{(\frac{1}{\mathrm{V_{f.u}}}-\frac{1}{n_{\mathrm{ox}}})} + 133.250 \frac{n_d^3}{(\mathrm{V_{f.u}})^3} + 11.349$ | 0.241 | 0.773 |
| 3 | $0.096\, n_{\mathrm{ox}}^2\, e^{-n_d} - 3.298\, e^{\frac{1}{\mathrm{V_{f.u}}}} \ln(1+n_{\mathrm{ox}}) + 119.859 \frac{n_d^3}{(\mathrm{V_{f.u}})^3} + 8.801$ | 0.243 | 0.774 |
| | **Model equations (set 2)** | **RMSE (eV)** | $R^2$ |
| 1 | $0.626 \frac{\Delta_{d-p}}{\ln(1+W)} - 1.963\sqrt{\frac{n_d}{W}} + 0.275 \frac{{n_d}^2}{\ln(1+\mathrm{V_{f.u}})} + 3.559$ | 0.207 | 0.813 |
| 2 | $0.321\, (\Delta_{d-p}) e^{\frac{1}{W}} - 1.604\sqrt{\frac{n_d}{W}} + 0.262 \frac{{n_d}^2}{\ln(1+\mathrm{V_{f.u}})} + 3.204$ | 0.210 | 0.812 |
| 3 | $0.022\, (\Delta_{d-p})^2 n_d - 2.235\, e^{(\frac{1}{W}-\frac{1}{n_d})} + 0.302 \frac{{n_d}^2}{\sqrt{\mathrm{V_{f.u}}}} + 4.732$ | 0.225 | 0.812 |

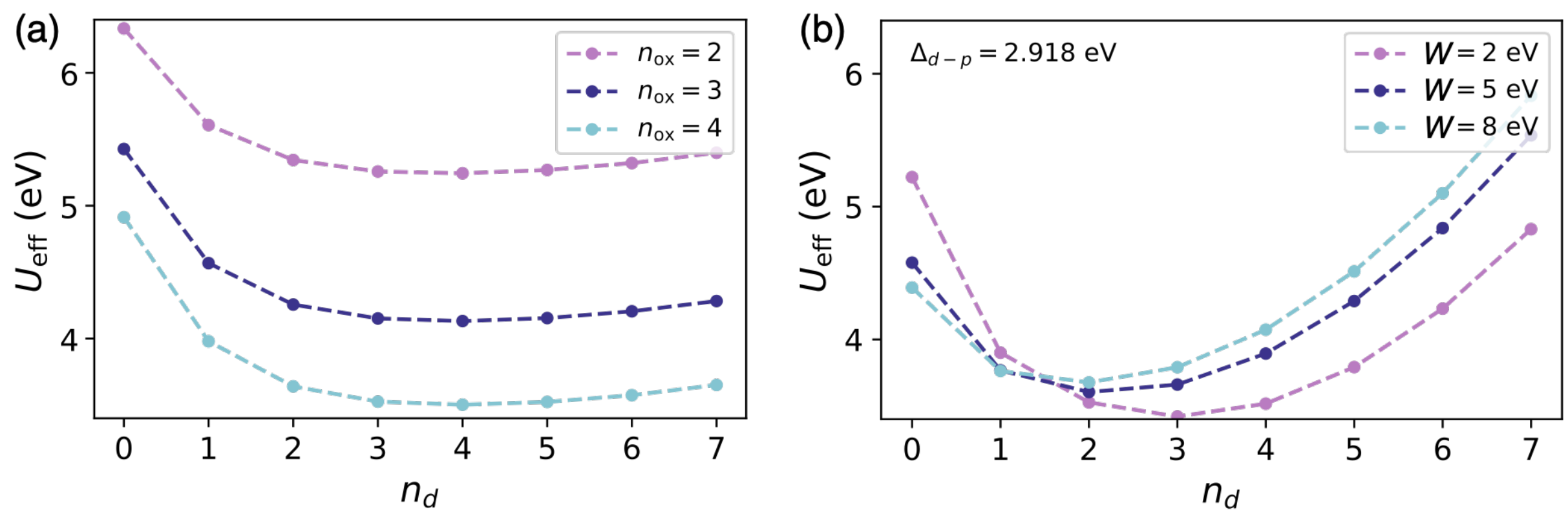

**Figure S1.** Dependence of $U_{\mathrm{eff}}$ on the number of $d$ electrons of TM ion, $n_d$ (a) for set 1 and (b) set 2, both at fixed volume $\mathrm{V_{f.u}} = 59.753$ Å$^3$. For set 2, $\Delta_{d\text{-}p}$ was fixed at its average value of 2.918 eV.

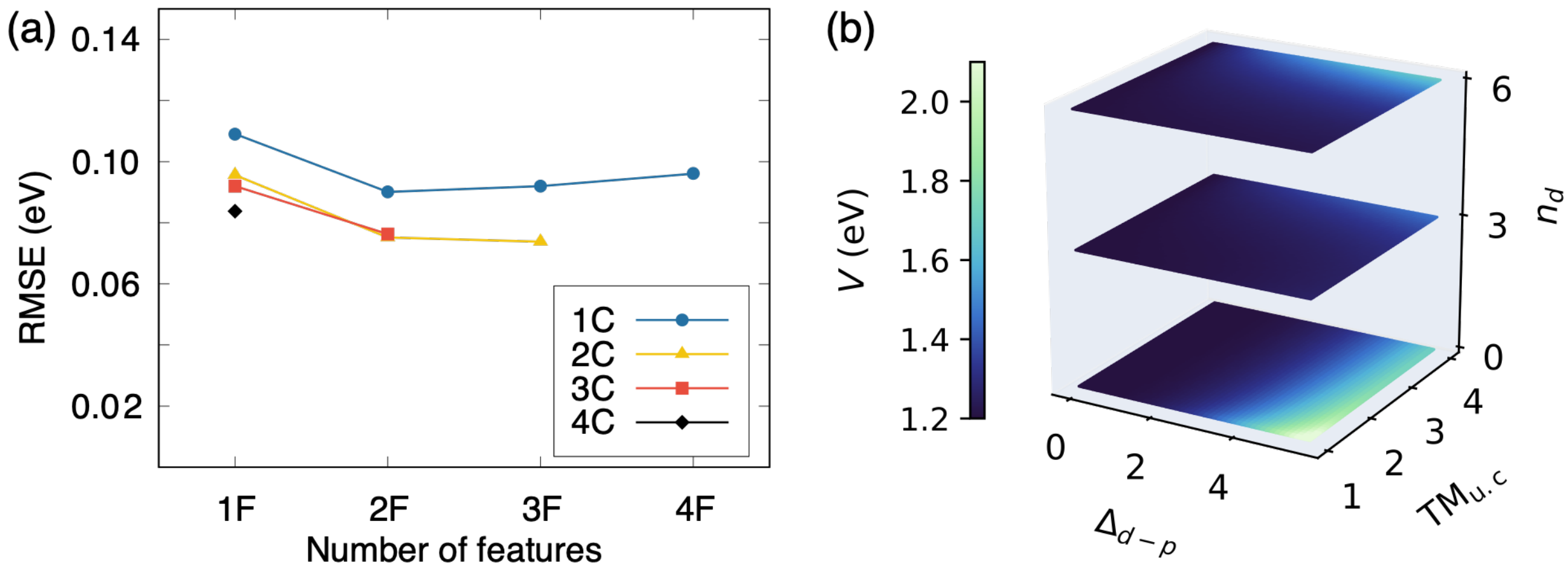

**Figure S2.** BFS results for $V$ using feature set 2 ($\Delta_{d\text{-}p}$, $W$, $n_d$, and $TM_{u.c}$) (a) Model performance depending on the number of features. (b) Three-dimensional (3D) surfaces of the best-performing BFS model.

**Table S5.** Equations of the three best-performing BFS models for predicting $V$.

| | 3C2F Model equations (set 1) | RMSE (eV) | $R^2$ |
|---|---|---|---|
| 1 | $0.919\,\frac{e^{-n_d}}{\sqrt{N_{u.c}}} + 1.673\times10^{-3}\,\frac{(N_{u.c})^3}{V_{f.u}} + 8.342\times10^{9}\,e^{-V_{f.u}}(N_{u.c})^3 + 1.193$ | 0.071 | 0.763 |
| 2 | $0.710\,\frac{e^{-n_d}}{\ln(1+N_{u.c})} + 1.670\times10^{-3}\,\frac{(N_{u.c})^3}{V_{f.u}} + 8.352\times10^{9}\,e^{-V_{f.u}}(N_{u.c})^3 + 1.193$ | 0.071 | 0.759 |
| 3 | $2.234\,\frac{e^{-n_d}}{N_{u.c}} + 1.674\times10^{-3}\,\frac{(N_{u.c})^3}{V_{f.u}} + 8.252\times10^{9}\,e^{-V_{f.u}}(N_{u.c})^3 + 1.197$ | 0.072 | 0.765 |
| | **2C3F Model equations (set 2)** | **RMSE (eV)** | **$R^2$** |
| 1 | $2.363\times10^{-3}\,(\Delta_{d-p})^3 e^{(\frac{1}{TM_{u.c}}-n_d)} + 8.392\times10^{-4}\,n_d\ln(1+\Delta_{d-p})\,e^{TM_{u.c}} + 1.197$ | 0.074 | 0.697 |
| 2 | $4.437\times10^{-3}\,\frac{(\Delta_{d-p})^3 e^{-n_d}}{\ln(1+TM_{u.c})} + 8.400\times10^{-4}\,n_d\ln(1+\Delta_{d-p})\,e^{TM_{u.c}} + 1.197$ | 0.074 | 0.696 |
| 3 | $7.689\times10^{-3}\,\frac{e^{(\Delta_{d-p}-n_d)}}{TM_{u.c}} + 8.363\times10^{-4}\,n_d\ln(1+\Delta_{d-p})\,e^{TM_{u.c}} + 1.199$ | 0.076 | 0.695 |

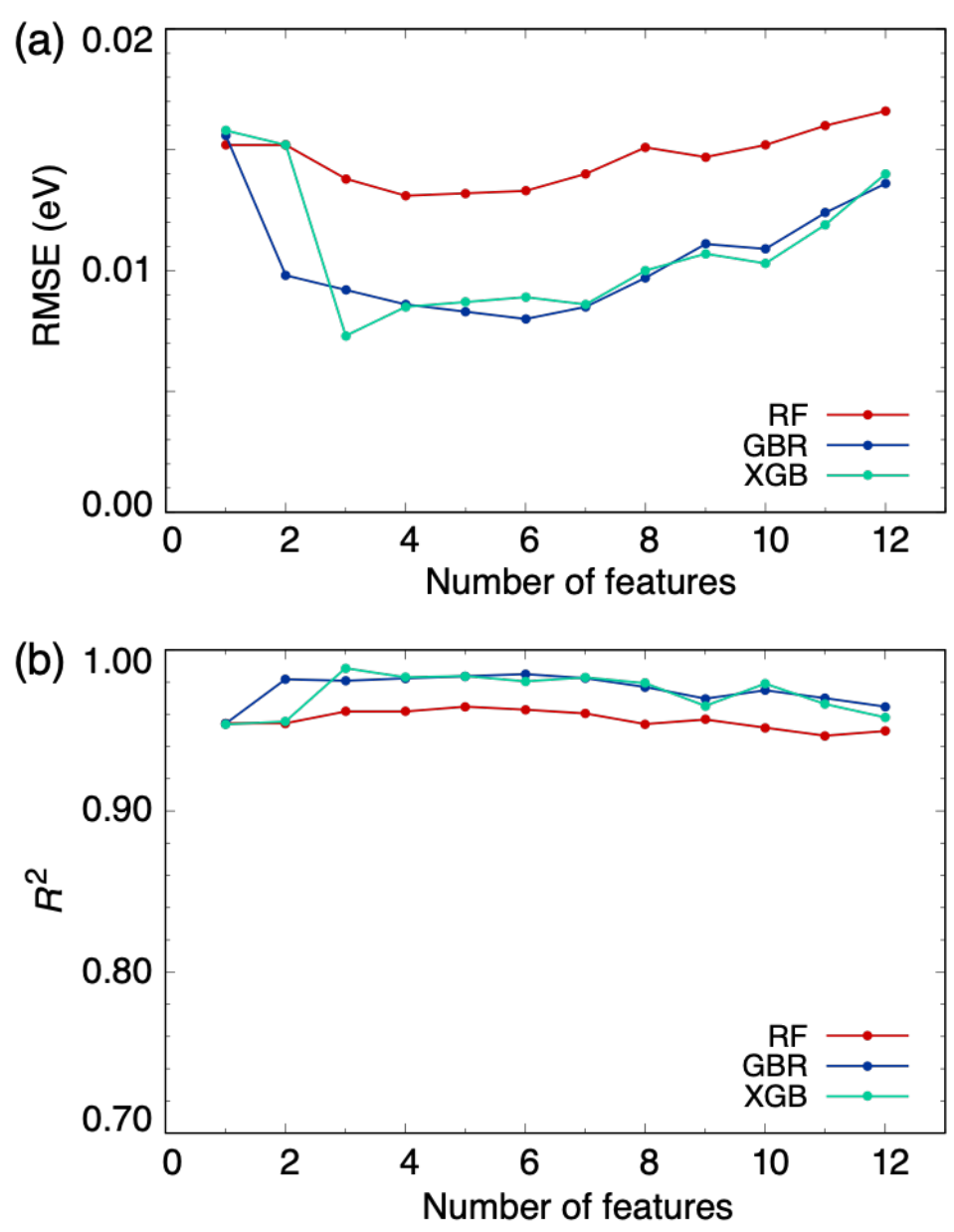

**Figure S3.** Prediction performance of ensemble-learning models for *J*. (a) RMSE and (b) $R^2$ values as functions of the number of selected features.

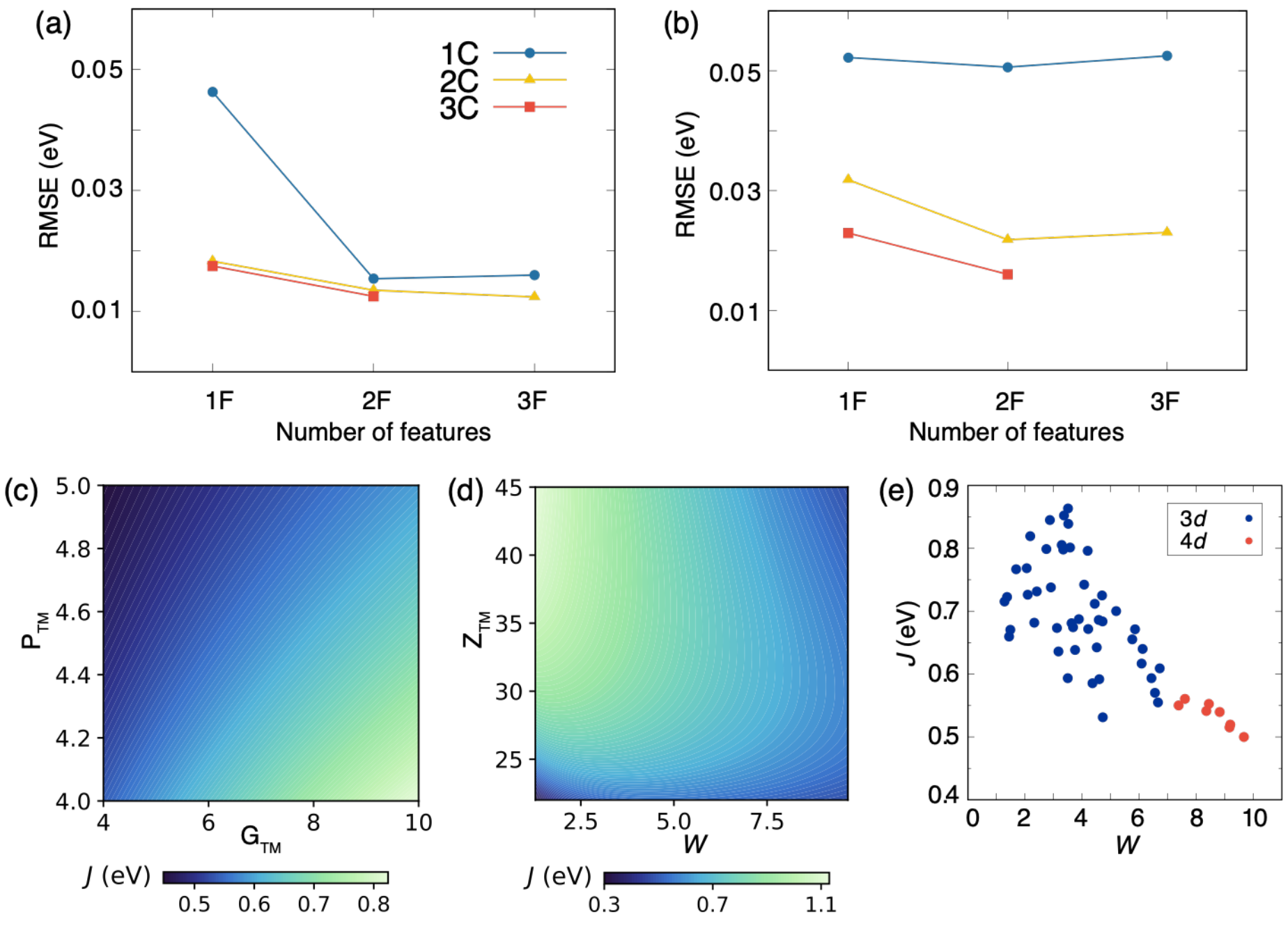

**Figure S4.** Prediction performance of BFS models for *J*. (a,b) RMSE as a function of the number of features using feature sets 1 and 2, respectively. (c,d) Two-dimensional plots of the representative BFS model for set 1 (1C2F) and set 2 (2C2F), respectively. (e) Relationship between *J* and *W*.

**Table S6.** Equations of representative BFS models for predicting *J*.

| | 1C2F Model equations (set 1) | RMSE (eV) | $R^2$ |
|---|---|---|---|
| 1 | $3.001\frac{\mathrm{G_{TM}}}{\mathrm{P_{TM}^3}}+0.354$ | 0.015 | 0.954 |
| 2 | $0.197\,\mathrm{G_{TM}^2}e^{-\mathrm{P_{TM}}}+0.488$ | 0.020 | 0.941 |
| 3 | $2.256\,\mathrm{G_{TM}}e^{-\mathrm{P_{TM}}}+0.399$ | 0.020 | 0.935 |
| | **2C2F Model equations (set 2)** | **RMSE (eV)** | **$R^2$** |
| 1 | $-0.055\sqrt{W\mathrm{Z_{TM}}}-1.287\times10^4\frac{1}{\sqrt{W}\mathrm{Z_{TM}^3}}+1.666$ | 0.022 | 0.929 |
| 2 | $-815.016\frac{1}{\sqrt{W}\mathrm{Z_{TM}^2}}-0.099\ln(1+W)\sqrt{\mathrm{Z_{TM}}}+2.148$ | 0.023 | 0.922 |
| 3 | $-0.073\ln(1+W)\sqrt{\mathrm{Z_{TM}}}-9.933\times10^3\frac{1}{\ln(1+W)\mathrm{Z_{TM}^3}})+1.678$ | 0.024 | 0.918 |

**Table S7.** List of transition-metal oxide compounds, the corresponding [$E_{min}$, $E_{max}$] energy window (in eV) used for calculating the TM-*d* and O-*p* band centers, and the disentanglement energy window $\mathbb{W}$ (in eV) used for Wannier-function construction. All energy values are given relative to the Fermi level ($E_F$ = 0 eV).

| Structure type | List | [$E_{min}$, $E_{max}$] | $\mathbb{W}$ |
|---|---|---|---|
| Perovskite | $SrTiO_3$ | [-8.0, 9.0] | Ref. 1 |
| | $SrVO_3$ | [-8.0, 6.0] | Ref. 1 |
| | $SrCrO_3$ | [-8.0, 5.0] | Ref. 1 |
| | $SrMnO_3$ | [-8.0, 5.0] | Ref. 1 |
| | $SrFeO_3$ | [-8.0, 4.0] | Ref. 1 |
| | $SrCoO_3$ | [-8.0, 4.0] | Ref. 1 |
| | $SrMoO_3$ | [-9.0, 8.0] | [-7.82;7.88] |
| | $SrTcO_3$ | [-9.0, 7.0] | [-8.32;6.48] |
| | $SrRuO_3$ | [-9.0, 7.0] | [-7.99;5.81] |
| | $SrRhO_3$ | [-9.0, 6.0] | [-7.67;5.43] |
| Olivine | $LiMnPO_4$ | [-11.0, 2.0] | [-10.92;1.08] |
| | $LiFePO_4$ | [-11.0, 2.0] | [-10.64;0.76] |
| | $NaMnPO_4$ | [-11.0, 2.0] | [-10.72;1.08] |
| | $NaFePO_4$ | [-11.0, 2.0] | [-10.39;0.81] |
| | $NaCoPO_4$ | [-11.0, 2.0] | [-10.50;0.80] |
| | $MnPO_4$ | [-11.0, 3.0] | [-10.03;1.87] |
| | $FePO_4$ | [-11.0, 3.0] | [-9.64;1.56] |
| | $CoPO_4$ | [-11.0, 3.0] | [-9.65;1.25] |
| | $NiPO_4$ | [-11.0, 3.0] | [-9.27;1.03] |
| Spinel | $LiCr_2O_4$ | [-9.0, 4.0] | [-8.02;3.08] |
| | $LiMn_2O_4$ | [-9.0, 4.0] | [-7.08;2.62] |
| | $LiFe_2O_4$ | [-9.0, 4.0] | [-7.16;2.54] |
| | $LiCo_2O_4$ | [-9.0, 4.0] | [-6.46;2.14] |
| | $LiNi_2O_4$ | [-9.0, 4.0] | [-7.46;1.04] |
| | $Li_2Ti_2O_4$ | [-9.0, 4.0] | [-8.37;3.63] |
| | $Li_2V_2O_4$ | [-9.0, 4.0] | [-8.26;2.94] |
| | $Li_2Mn_2O_4$ | [-9.0, 4.0] | [-7.22;2.58] |
| | $Li_2Co_2O_4$ | [-9.0, 4.0] | [-7.10;1.30] |
| | $Mn_2O_4$ | [-9.0, 4.0] | [-7.09;3.20] |
| Layered | $LiVO_2$ | [-9.0, 4.0] | [-8.58;3.52] |
| | $LiCrO_2$ | [-9.0, 4.0] | [-8.10;2.60] |
| | $LiMnO_2$ | [-9.0, 4.0] | [-7.69;2.71] |
| | $LiFeO_2$ | [-9.0, 4.0] | [-6.92;1.88] |
| | $LiCoO_2$ | [-9.0, 4.0] | [-8.12;1.48] |
| | $LiNiO_2$ | [-9.0, 4.0] | [-7.98;0.72] |
| | $NaVO_2$ | [-8.0, 4.0] | [-7.19;2.61] |
| | $NaCrO_2$ | [-8.0, 4.0] | [-6.98;2.12] |
| | $NaMnO_2$ | [-8.0, 4.0] | [-6.56;2.14] |
| | $NaFeO_2$ | [-8.0, 4.0] | [-5.98;1.42] |
| | $NaCoO_2$ | [-8.0, 4.0] | [-6.90;1.10] |
| | $NaNiO_2$ | [-8.0, 4.0] | [-6.71;0.69] |
| | $MnO_2$ | [-8.0, 4.0] | [-6.69;3.01] |
| | $CoO_2$ | [-8.0, 4.0] | [-6.76;2.84] |
| | $NiO_2$ | [-8.0, 4.0] | [-7.13;1.77] |
| Layered perovskite | $Sr_2TiO_4$ | [-8.0, 8.0] | [-5.33;7.95] |
| | $Sr_2VO_4$ | [-8.0, 5.5] | [-7.08;5.72] |
| | $Sr_2CrO_4$ | [-8.0, 5.0] | [-7.03;4.97] |
| | $Sr_2MnO_4$ | [-8.0, 5.0] | [-6.93;4.47] |
| | $Sr_2FeO_4$ | [-8.0, 4.0] | [-6.42;3.88] |
| | $Sr_2CoO_4$ | [-8.0, 4.0] | [-6.39;3.71] |
| | $Sr_2MoO_4$ | [-10.0, 9.0] | [-8.36;8.34] |
| | $Sr_2TcO_4$ | [-10.0, 9.0] | [-8.16;7.34] |
| | $Sr_2RhO_4$ | [-10.0, 9.0] | [-8.07;6.83] |
| | $Sr_2RuO_4$ | [-10.0, 9.0] | [-7.87;6.13] |

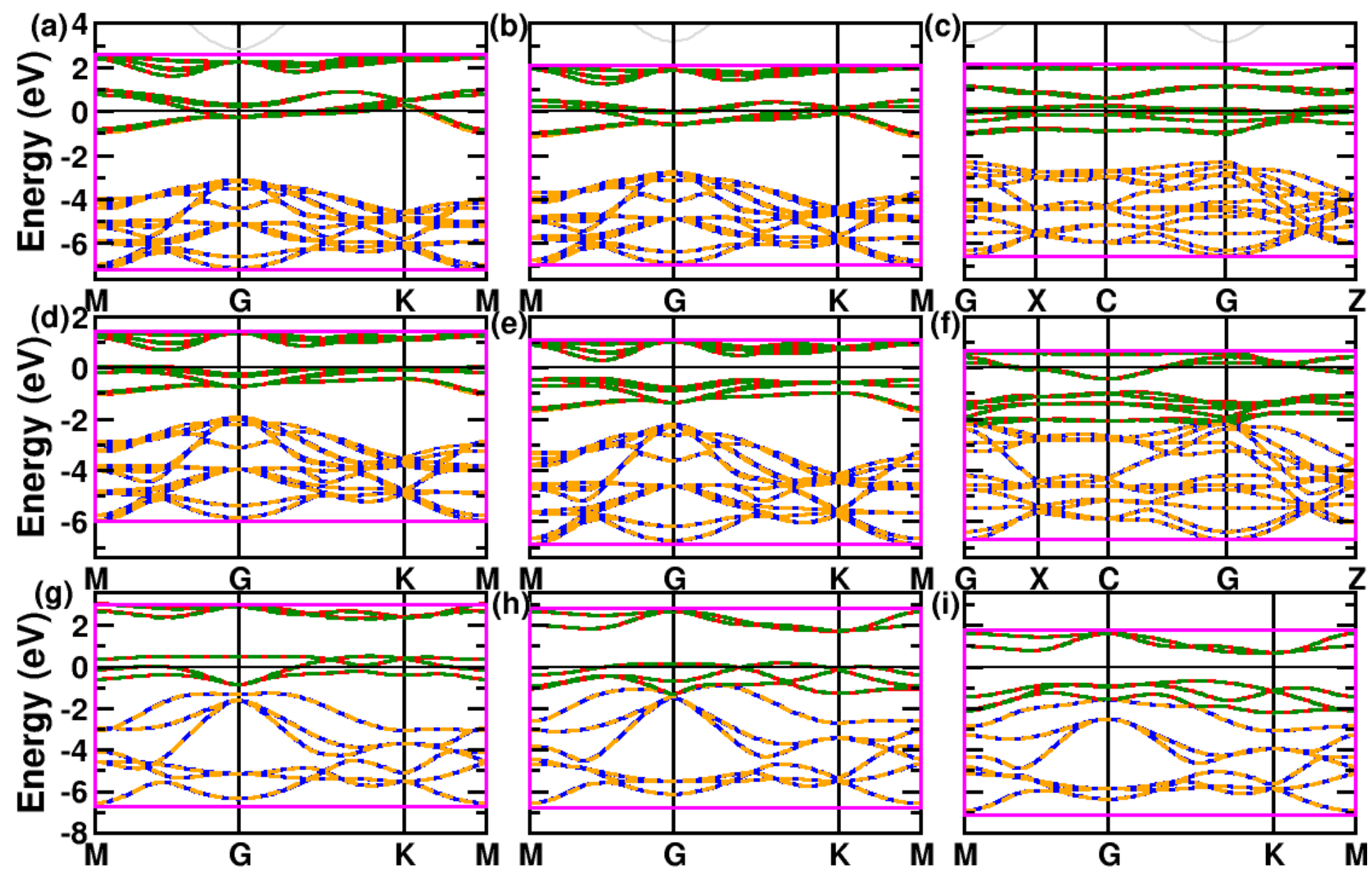

**Figure S5.** Nonmagnetic band structures of layered materials together with their Wannier-interpolated bands. The solid blue (red) and dashed orange (green) lines represent the DFT and Wannier-interpolated bands of O-*p* (TM-*d*) bands, respectively. The disentanglement energy windows for constructing the Wannier functions are represented by solid magenta lines. **(a)** $NaVO_2$, **(b)** $NaCrO_2$, **(c)** $NaMnO_2$, **(d)** $NaFeO_2$, **(e)** $NaCoO_2$, **(f)** $NaNiO_2$, **(g)** $MnO_2$, **(h)** $CoO_2$, and **(i)** $NiO_2$.

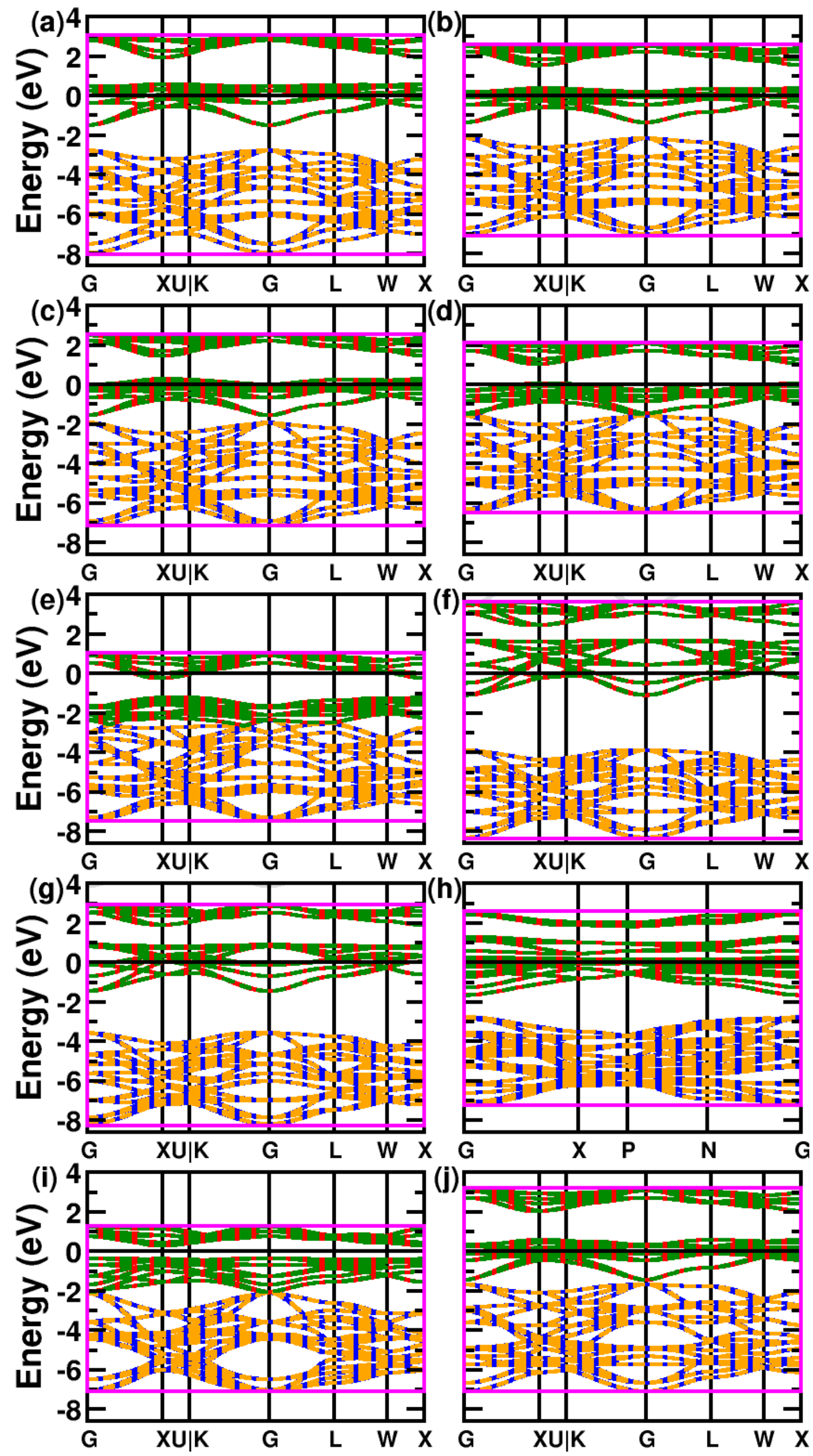

**Figure S6.** Nonmagnetic band structures of spinel materials together with their Wannier-interpolated bands. Line colors and styles are identical to those used in Figure S5. **(a)** $LiCr_2O_4$, **(b)** $LiMn_2O_4$, **(c)** $LiFe_2O_4$, **(d)** $LiCo_2O_4$, **(e)** $LiNi_2O_4$, **(f)** $Li_2Ti_2O_4$, **(g)** $Li_2V_2O_4$, **(h)** $Li_2Mn_2O_4$, (**i**) $Li_2Co_2O_4$, and **(j)** $Mn_2O_4$.

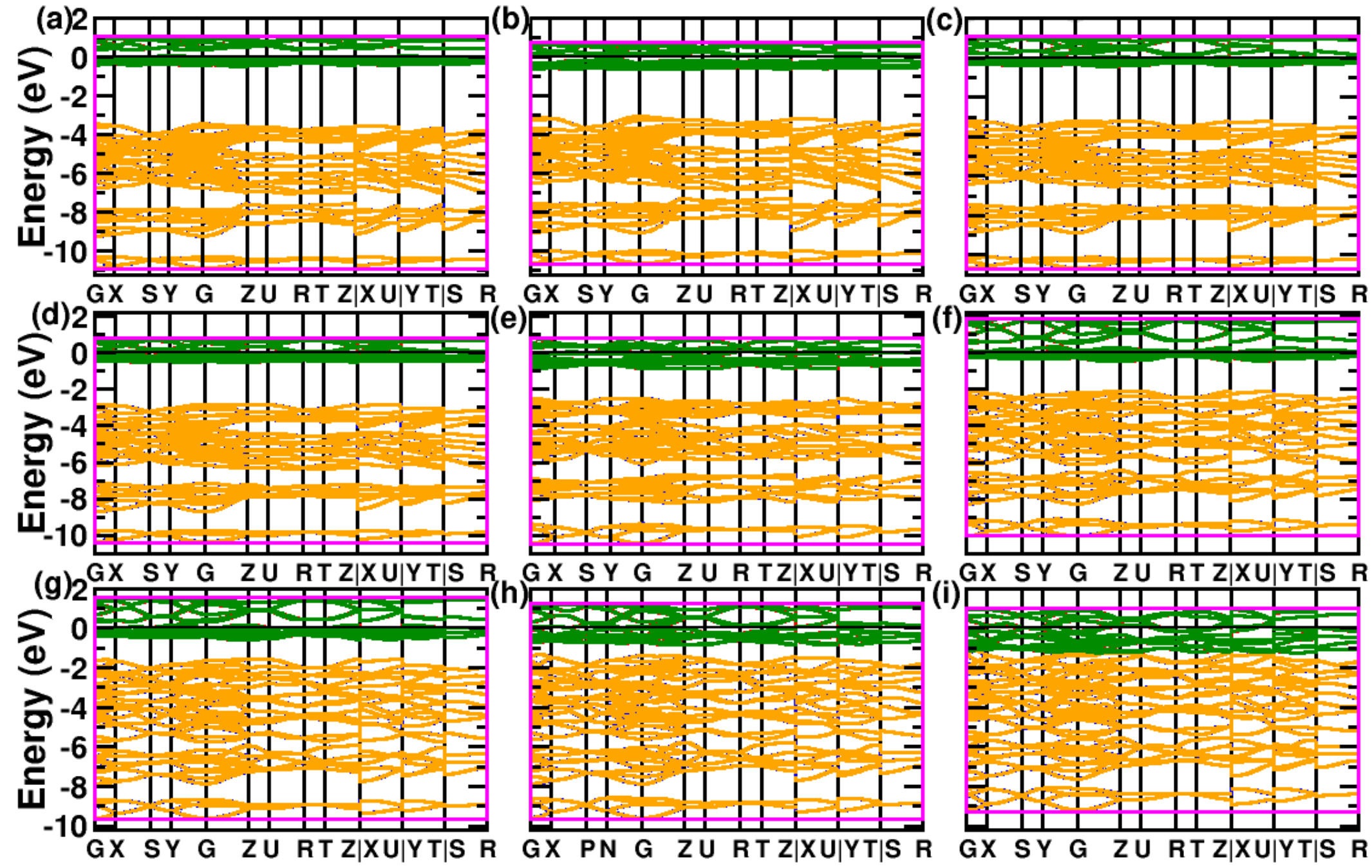

**Figure S7.** Nonmagnetic band structures of olivine materials together with their Wannier-interpolated bands. Line colors and styles are identical to those used in Figure S5. **(a)** $LiMnPO_4$, **(b)** $LiFePO_4$, **(c)** $NaMnPO_4$, **(d)** $NaFePO_4$, **(e)** $NaCoPO_4$, **(f)** $MnPO_4$, **(g)** $FePO_4$, **(h)** $CoPO_4$, and **(i)** $NiPO_4$.

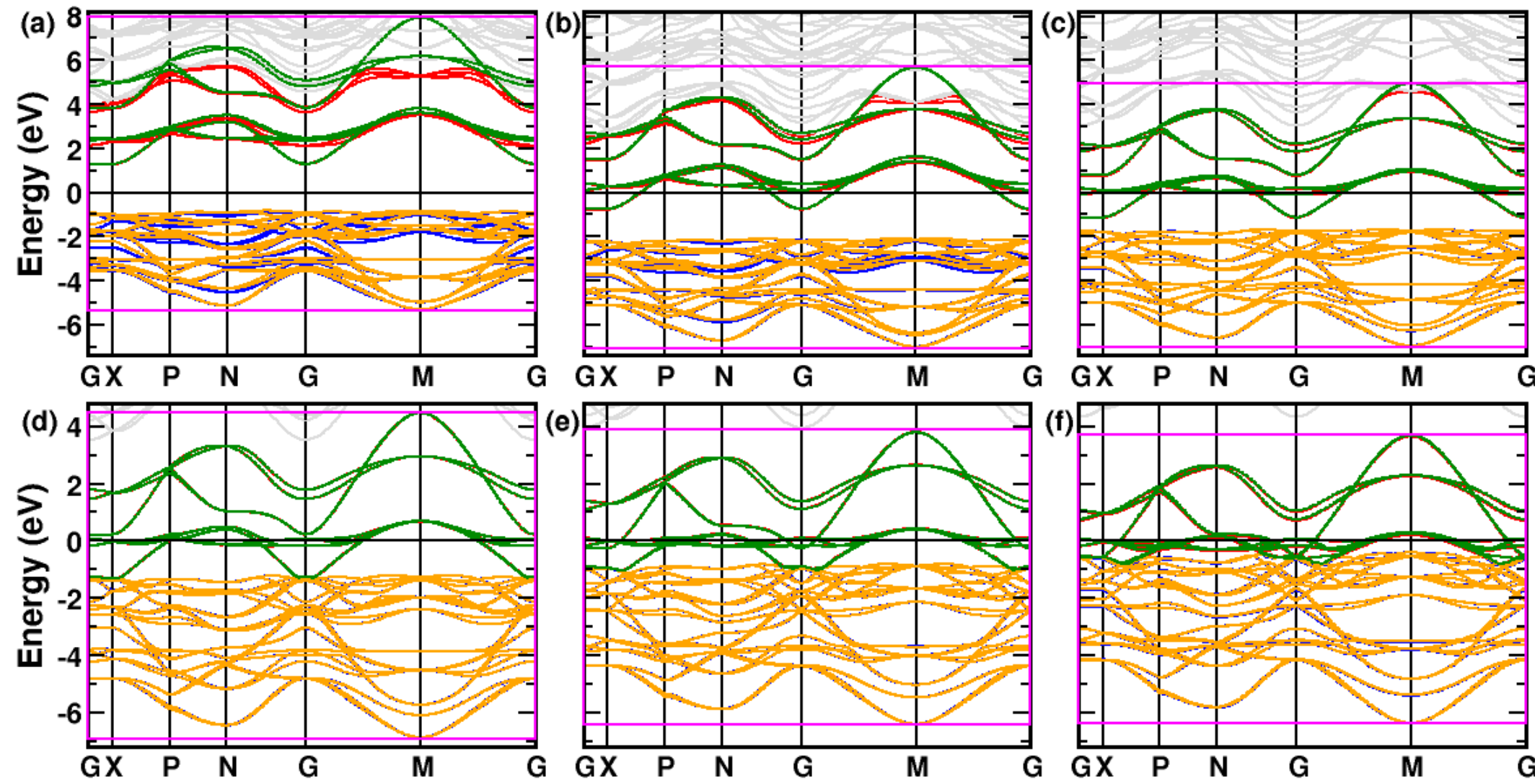

**Figure S8.** Nonmagnetic band structures of $Sr_2MO_4$ (M = 3*d*) layered perovskite materials together with their Wannier-interpolated bands. Line colors and styles are identical to those used in Figure S5. **(a)** $Sr_2TiO_4$, **(b)** $Sr_2VO_4$, **(c)** $Sr_2CrO_4$, **(d)** $Sr_2MnO_4$, **(e)** $Sr_2FeO_4$, and **(f)** $Sr_2CoO_4$.

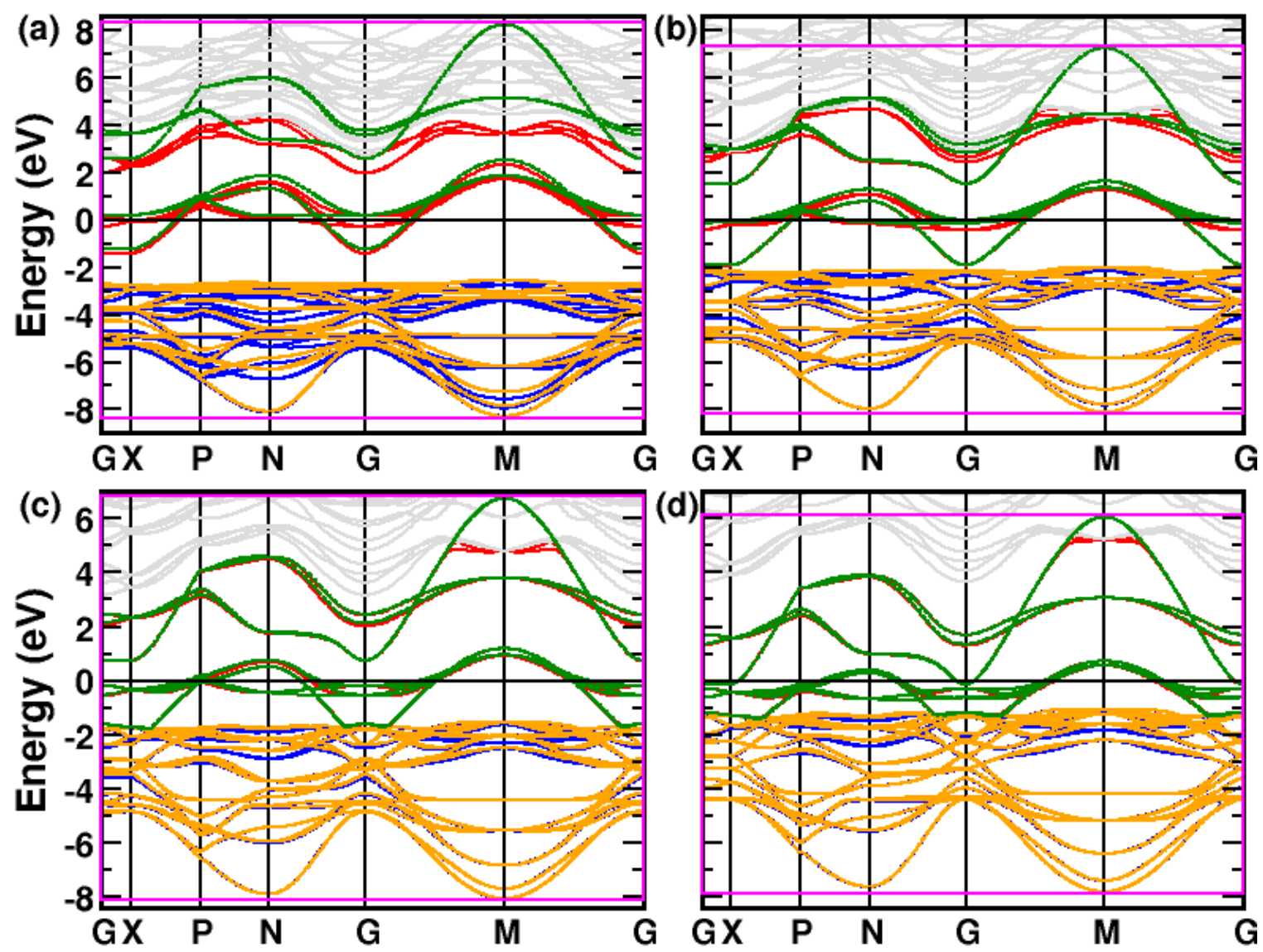

**Figure S9.** Nonmagnetic band structures of $Sr_2MO_4$ (M = 4*d*) layered perovskite materials together with their Wannier-interpolated bands. Line colors and styles are identical to those used in Figure S5. **(a)** $Sr_2MoO_4$, **(b)** $Sr_2TcO_4$, **(c)** $Sr_2RuO_4$, and **(d)** $Sr_2RhO_4$.

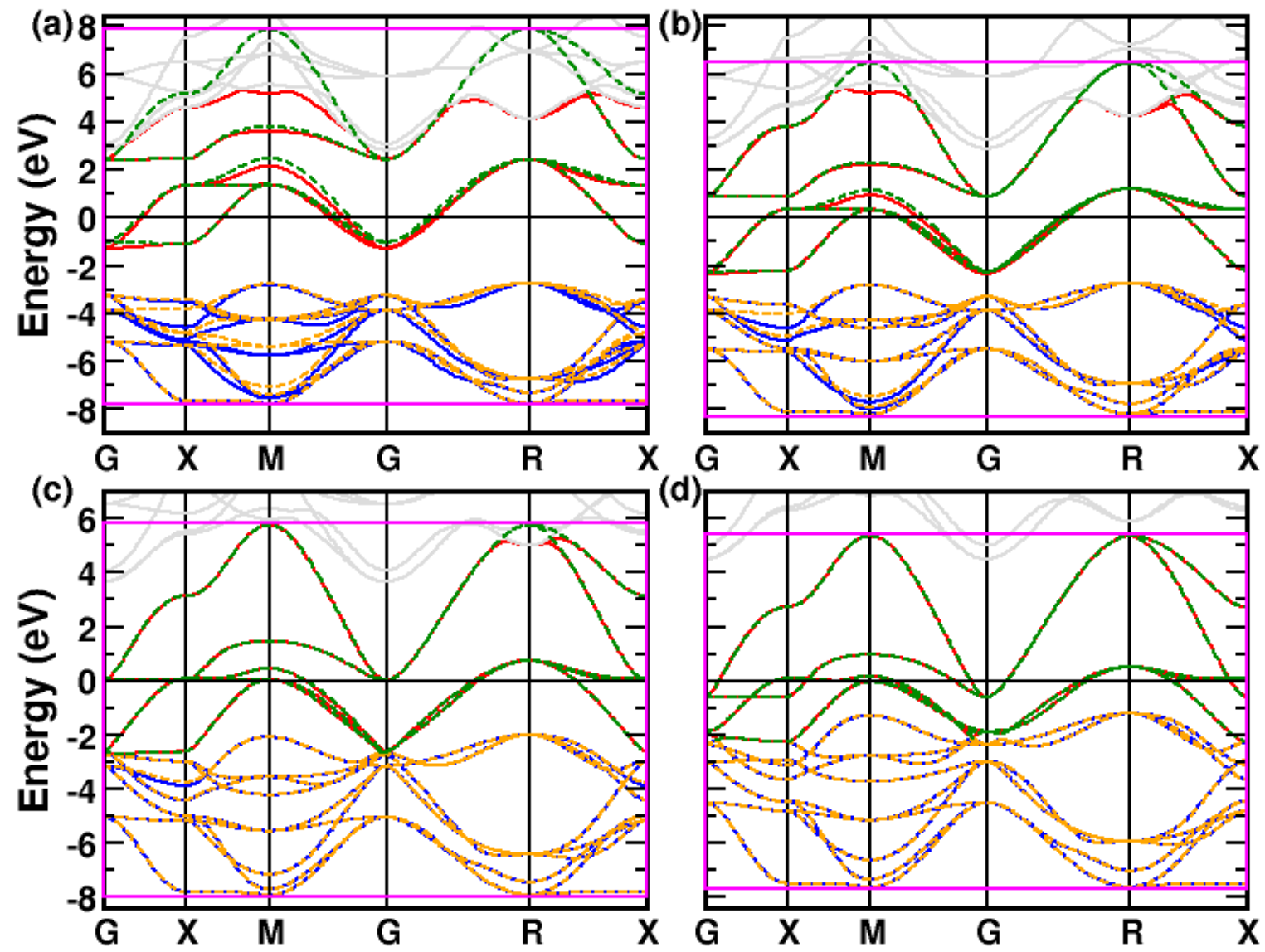

**Figure S10.** Nonmagnetic band structures of $SrMO_3$ (M = 4*d*) perovskite materials together with their Wannier-interpolated bands. Line colors and styles are identical to those used in Figure S5. **(a)** $SrMoO_3$, **(b)** $SrTcO_3$, **(c)** $SrRuO_3$, and **(d)** $SrRhO_3$.